\documentclass[prd,aps,twocolumn,a4paper,showkeys,nofootinbib]{revtex4-1}

\usepackage{graphicx,psfrag}
\usepackage{mathrsfs}
\usepackage{amsmath,amsfonts,amssymb}
\usepackage{comment}

\usepackage{enumitem}

\usepackage{xcolor}
\definecolor{linkcolor}{rgb}{0.1,0.7,0.3}
\usepackage[unicode, colorlinks=true, linkcolor=linkcolor, citecolor=linkcolor, filecolor=linkcolor, urlcolor=linkcolor, linktocpage, breaklinks]{hyperref}

\usepackage{longtable}
\usepackage{multirow}

\usepackage[utf8]{inputenc}
\usepackage[T1]{fontenc}

\usepackage[printonlyused]{acronym}

\allowdisplaybreaks

\newcommand{\be}{\begin{equation}}
\newcommand{\ee}{\end{equation}}
\newcommand{\bea}{\begin{eqnarray}}
\newcommand{\eea}{\end{eqnarray}}
\newcommand{\bel}{\begin{align}}
\newcommand{\eel}{\end{align}}

\newcommand{\orcid}[1]{\href{https://orcid.org/#1}{
\includegraphics[width=10pt]{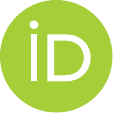}
}}

\def\lm{{\ell m}}

\def\F{{\cal F}}
\def\tmatch{t_\lm^{\rm match}}
\def\eg{\textit{e.g.}}
\def\ie{\textit{i.e.}}

\usepackage{pifont} 

\def\DALI{\texttt{TEOBResumS-Dal\'i}}
\def\GIOTTO{\texttt{TEOBResumS-GIOTTO}}

\def\Surr{\texttt{NRHybSur3dq8}}
\def\SurrCCE{\texttt{NRHybSur3dq8{\_}CCE}}
\def\RWZ{\texttt{RWZHyp}}
\def\Teuk{\texttt{Teukode}}
\def\ta{\tilde{a}_0}
\def\tA22{t_{A_{22}}^{\rm peak}}
\def\tNQC{t_\lm^{\rm NQC}}

\usepackage{color}
\definecolor{cyan}{rgb}{0,0.9,0.9}
\definecolor{orange}{rgb}{0.9,0.5,0}
\definecolor{magenta}{rgb}{1,0,1}
\definecolor{purple}{rgb}{0.8,0.4,0.8}
\definecolor{gray}{rgb}{0.8242,0.8242,0.8242}

\newcommand{\HT}[1]{{\cal H}\left[{#1}\right]}
\newcommand{\lettersection}[1]{\vspace{0.1cm}{\textbf{\textit{#1}}} --}

\begin{document}

\title{Real modes and null memory contributions in effective-one-body models}

\author{Simone \surname{Albanesi} \orcid{0000-0001-7345-4415}}
\email{simone.albanesi@uni-jena.de}
\affiliation{Theoretisch-Physikalisches Institut, Friedrich-Schiller-Universit{\"a}t Jena, 07743, Jena, Germany}
\affiliation{INFN sezione di Torino, Torino, 10125, Italy}

\date{\today}

\begin{abstract}
We introduce a novel approach to describe real-valued $m=0$ modes from inspiral to merger 
and ringdown in effective-one-body models, including both oscillatory and null memory contributions.
A crucial aspect of the modelization of the oscillatory part is the complexification of the real modes
via a Hilbert transform. This procedure allows for an accurate description of the merger-ringdown 
waveform by applying standard approaches employed for the complex $m>0$ modes, which include 
source-driven effects. The physical signal is then recovered by solely considering the real part.
We apply this method in the extreme-mass-ratio regime, 
considering particle-driven linear gravitational perturbations in Schwarzschild and Kerr spacetimes.
We then extend our description to spin-aligned, quasi-circular, comparable-mass binaries 
providing hierarchical fits incorporating the test-mass limit.
The post-merger waveform is then matched with an inspiral effective-one-body waveform.
By adopting \GIOTTO{} as our baseline,  
we also include the displacement memory in the (2,0) mode through Bondi–van der Burg-Metzner–Sachs
balance laws, thus providing a complete effective-one-body model incorporating both oscillatory 
and null memory effects. The accuracy of this model is validated against 
the hybrid numerical relativity surrogate \SurrCCE{},
finding, for the quadrupole of the equal mass nonspinning case, a LIGO noise-weighted mismatch
of $\bar{{\cal F}} = 6\times 10^{-4}$ at $50 M_\odot$ 
for the inclination that maximizes the contribution of the (2,0) mode.
\end{abstract}

\maketitle

\acrodef{GW}{gravitational wave}
\acrodef{PN}{post-Newtonian}
\acrodef{NR}{numerical relativity}
\acrodef{QNM}{quasi-normal-mode}
\acrodef{NQC}{next-to-quasicircular}
\acrodef{BMS}{Bondi-van der Burg-Metzner-Sachs}
\acrodef{BBH}{binary black hole}
\acrodef{CCE}{Cauchy characteristic extraction}
\acrodef{IMR}{inspiral-merger-ringdown}
\acrodef{LVK}{LIGO-Virgo-KAGRA}
\acrodef{SM}{Supplemental Material}
\acrodef{EOB}{effective-one-body}

\lettersection{Introduction}
%
With the detections from the \ac{LVK} network~\cite{LIGOScientific:2014pky,VIRGO:2014yos}, \ac{GW} astronomy has 
now become a standard tool for exploring the 
Universe~\cite{LIGOScientific:2018mvr,LIGOScientific:2020ibl,LIGOScientific:2021djp}.
Waveform models for compact binaries are central to this field, both for detection and data analysis, 
making it essential to provide accurate \ac{GW} models.
Moreover, the advent of next-generation detectors, such as Einstein 
Telescope~\cite{Punturo:2010zz,Maggiore:2019uih}, LISA~\cite{LISA:2017pwj}, 
and Cosmic Explorer~\cite{Reitze:2019iox}, requires even further advancements in waveform modeling, 
not only in accuracy~\cite{Purrer:2019jcp}, but also in capturing all the physically relevant phenomena~\cite{Gupta:2024gun}. 
\begin{figure}[t]
  \centering 
    \includegraphics[width=0.48\textwidth]{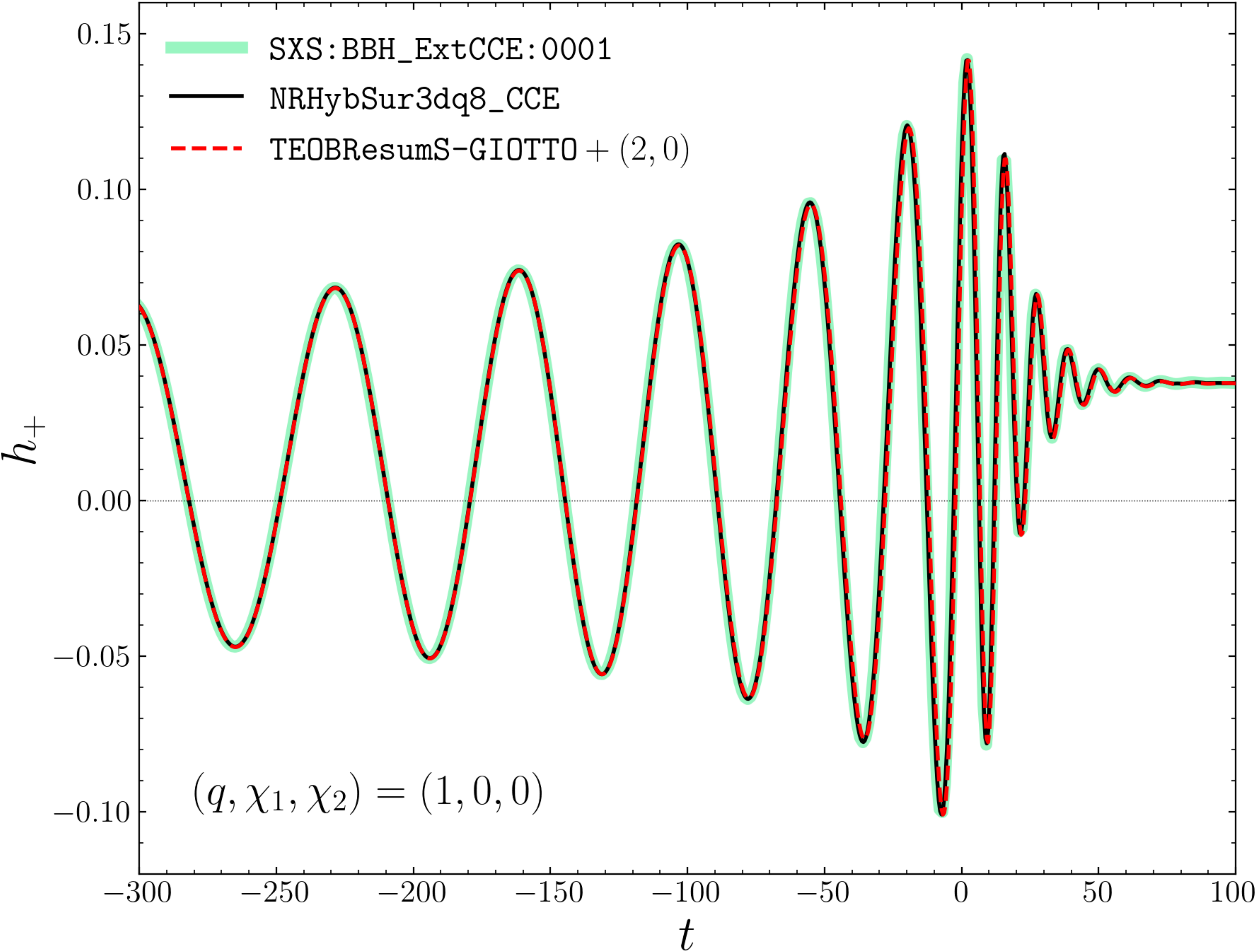}
    \caption{Quadrupolar strains for the equal mass nonspinning case and
    observational direction $(\Theta,\Phi)=(\frac{\pi}{2},0)$. We consider 
    the simulation \texttt{SXS:BBH{\_}ExtCCE:0001}~\cite{Mitman:2021xkq} (thick green),
    the surrogate \SurrCCE{}~\cite{Yoo:2023spi} (solid black), and     
    the EOB model \GIOTTO{}~\cite{Nagar:2023zxh} completed with the 
    (2,0) mode as discussed in this work (dashed red).
    }
  \label{fig:eobnr_strain}
\end{figure}
The \ac{EOB} approach~\cite{Buonanno:1998gg,Buonanno:2000ef,Damour:2001tu}
is capable to accurately model \acp{GW} from compact binaries, 
once calibrated and complemented with \ac{NR} data~\cite{Buonanno:2006ui,Damour:2007xr,Damour:2007vq}.
This approach does not only describe standard quasi-circular \acp{BBH}~\cite{Nagar:2023zxh,Pompili:2023tna}, 
but can be also employed to model binaries in presence of
matter~\cite{Damour:2009wj,Damour:2012yf,Nagar:2018zoe,Dietrich:2018uni}, 
eccentricity~\cite{Chiaramello:2020ehz,Nagar:2020xsk,Nagar:2021gss,Khalil:2021txt,Placidi:2021rkh,Ramos-Buades:2021adz,Albanesi:2022xge,Nagar:2024dzj,Nagar:2024oyk},
and spin-precession~\cite{Buonanno:2005xu,Pan:2013rra,Akcay:2020qrj,Gamba:2021ydi,Khalil:2023kep,Ramos-Buades:2023ehm},
eventually jointly~\cite{Gonzalez:2022prs,Gamba:2022mgx,Gamba:2023mww,Gamba:2024cvy}.
However, state-of-the-art EOB models do not include $m=0$ modes in the co-precessing frame, despite them 
being particularly relevant when the radial acceleration is large, \eg{}, on eccentric orbits or during the plunge. 
In this work we devise a strategy to incorporate their description, from inspiral to merger and ringdown,
considering both oscillatory contributions and displacement memory effects.  
To build our model for the oscillatory part, we consider numerical results obtained by solving linear perturbations
sourced by test-particle dynamics~\cite{Regge:1957td,Zerilli:1970se,Nagar:2005ea,Martel:2005ir,Teukolsky:1973ha}
with \RWZ{}~\cite{Bernuzzi:2010ty,Bernuzzi:2011aj,Bernuzzi:2012ku} 
and \Teuk{}~\cite{Harms:2014dqa}, together with
waveforms obtained from the surrogate model \Surr{}~\cite{Varma:2018mmi}, which is built using \ac{NR} simulations of the SXS collaboration~\cite{Chu:2009md,Lovelace:2010ne,Lovelace:2011nu,Buchman:2012dw,Hemberger:2013hsa,Scheel:2014ina,Blackman:2015pia,Lovelace:2014twa,Mroue:2013xna,Kumar:2015tha,Chu:2015kft,Boyle:2019kee,SXS:catalog,Mitman:2021xkq}.
After having discussed some phenomenological aspects, we start by extending the post-merger model of 
Ref.~\cite{Damour:2014yha} to the real $m=0$ oscillatory modes through a complexification
achieved with a Hilbert transform~\cite{Hilbert:1912,Bracewell:2000}. We then provide 
a closed-form representation of the oscillatory post-merger for the (2,0) mode across the parameter space
of spin-aligned binaries with generic mass ratios 
by means of hierarchical fits. This post-merger model is then 
matched with an inspiral EOB waveform based on Ref.~\cite{Chiaramello:2020ehz},
using standard \ac{NQC} corrections~\cite{Damour:2007xr}. 

The (2,0) mode is also sourced by the outgoing gravitational radiation, thus
manifesting a nonlinear memory contribution~\cite{Christodoulou:1991cr,Thorne:1992sdb,Blanchet:1992br,Mitman:2024uss}. 
This effect, which is inherently related to the radiated energy flux~\cite{Thorne:1992sdb}, has been
studied for quasi-circular binaries in \ac{PN} theory~\cite{Favata:2008yd,Cunningham:2024dog}
and in minimal-waveform models, where an \ac{EOB} inspiral/plunge waveform is completed with a 
post-merger model purely based on a \ac{QNM} description~\cite{Favata:2009ii,Favata:2010zu}.
Null memory contributions for eccentric inspirals have also been studied within the PN framework~\cite{Favata:2011qi,Ebersold:2019kdc} and the EOB one~\cite{Grilli:2024lfh}.
These effects have been resolved in \ac{NR} simulations only recently~\cite{Pollney:2010hs,Mitman:2020pbt},
by using \ac{CCE}~\cite{Bishop:1996gt,Reisswig:2009rx,Moxon:2020gha,Moxon:2021gbv,deppe_2024_10967177}.
It should be noted that the strain, and memory effects in particular, 
are inherently linked to the \ac{BMS} group~\cite{Bondi:1962,Sachs:1962},
which is the symmetry group of asymptotically flat spacetimes at future null infinity
(Poincaré group plus supertranslations). Therefore, using \ac{BMS} balance laws,
null memory effects can also be added in post-processing to \ac{NR} simulations~\cite{Talbot:2018sgr,Mitman:2020bjf}.
The displacement memory in the (2,0) mode has been also included in some \ac{IMR} waveforms models~\cite{Favata:2009ii,Favata:2010zu,Cao:2016maz},
and in particular in \texttt{IMRPhenomTHM}~\cite{Rossello-Sastre:2024zlr}. 
Memory effects have been searched in \ac{LVK} data~\cite{Boersma:2020gxx,Hubner:2019sly}, 
and are potentially detectable with future 
observations~\cite{Grant:2022bla,Goncharov:2023woe,Gasparotto:2023fcg,Inchauspe:2024ibs,Xu:2024ybt}.  

We then apply our strategy employing \GIOTTO{}~\cite{Nagar:2023zxh} as a baseline, thus completing
this model by including the description of the (2,0) multipole for spin-aligned quasi-circular \acp{BBH} with generic mass ratio.
The oscillatory part is described
with an EOB waveform completed with the aforementioned Hilbert post-merger
model, while the null memory contribution is computed from \ac{BMS}
balance laws and matched, at low-frequencies, with the \ac{PN} result~\cite{Favata:2011qi,Ebersold:2019kdc,Cao:2016maz}.  
The model is validated against the hybrid \ac{NR}-surrogate \SurrCCE{}~\cite{Yoo:2023spi}, whose waveforms
are in the same BMS frame as the EOB ones, thus avoiding any coordinate ambiguity~\cite{Mitman:2021xkq,Mitman:2022kwt}.
A comparison for the quadrupolar strain with this surrogate and a \ac{NR} SXS 
simulation~\cite{Mitman:2021xkq} is anticipated
in Fig.~\ref{fig:eobnr_strain}. More details are discussed in the conclusions.

\lettersection{Conventions}
We use geometric units $G=c=1$. 
The \ac{GW} strain is decomposed
in multipoles on spin-weighted spherical harmonics as 
$h_+ - i h_\times = \sum_{\lm} {}_{-2}Y_{\lm}(\Theta, \Phi) R^{-1} h_\lm$, being
$R$ the distance of the observer. Multipoles
are further decomposed in amplitude and phase as $h_\lm = A_\lm e^{-i \phi_\lm}$.
The waveform frequency is computed as $\omega_{\lm}=\dot{\phi}_\lm$,
where the dot indicates a time-derivative.
If not specified otherwise, 
length-scales are normalized with the total rest mass of the system $M=m_1+m_2$, 
where $m_{1,2}$ are the masses of the two black holes ($m_1>m_2$).
The symmetric mass ratio is defined as $\nu = q/(1+q)^2$, where
$q=m_1/m_2$. 
The dimensionless Kerr spin is denoted as $\hat{a}$, with 
$\hat{a}\geq 0$ ($\hat{a}<0$) for prograde (retrograde) equatorial orbits.
For comparable-mass binaries, the dimensionless black hole spins are 
instead denoted as $\chi_{1,2}$. 
The time corresponding to the peak of the (2,2) amplitude
is indicated as $\tA22$. 

\lettersection{EOB dynamics and test-mass limit}
The standard \ac{EOB} approach consists in mapping the 2-body \ac{PN} dynamics
to the motion of a particle in an effective metric, which is a $\nu$-deformation
of the Schwarzschild or Kerr one~\cite{Buonanno:1998gg,Buonanno:2000ef,Damour:2001tu}.
The EOB Hamiltonian, together with a PN-resummed prescription
for the radiation reaction, provides Hamilton's equations that can be 
solved to obtain the effective dynamics, from which is then possible to analytically compute 
the multipolar waveform at infinity.
By setting to zero the $\nu$-corrections in the conservative sector and keeping 
leading-order contribution in the radiation reaction, we can evolve inspirals 
in Schwarzschild and Kerr spacetimes, thus obtaining a first-order description
of extreme-mass ratio binaries. However, note that higher-order corrections in the mass ratio
are needed to describe real astrophysical scenarios~\cite{Pound:2015tma}.
Given test-mass dynamics, we can numerically compute the linear-order 
gravitational perturbations sourced by the motion of the particle. This approach 
has been extensively used to gain insights into prescriptions
to use for binaries with generic mass-ratio~\cite{Damour:2007xr,Bernuzzi:2010xj,Taracchini:2014zpa,Albanesi:2023bgi}.
In this work, we start by considering the quasi-circular inspiral
of Ref.~\cite{Albanesi:2023bgi}, obtained by solving the Regge-Wheeler and Zerilli
equations with a test-particle source 
term~\cite{Regge:1957td,Zerilli:1970se,Nagar:2005ea,Martel:2005ir} using
the time-domain code \RWZ{}~\cite{Bernuzzi:2010ty,Bernuzzi:2011aj,Bernuzzi:2012ku}. 
We then explore the Kerr case, by solving the Teukolsky equation~\cite{Teukolsky:1973ha} 
for 11 quasi-circular configurations with \Teuk{}~\cite{Harms:2014dqa}. The test-mass simulations are
listed in Table~\ref{tab:sims_testmass} of the \ac{SM}.

\begin{figure}[t]
  \centering 
    \includegraphics[width=0.48\textwidth]{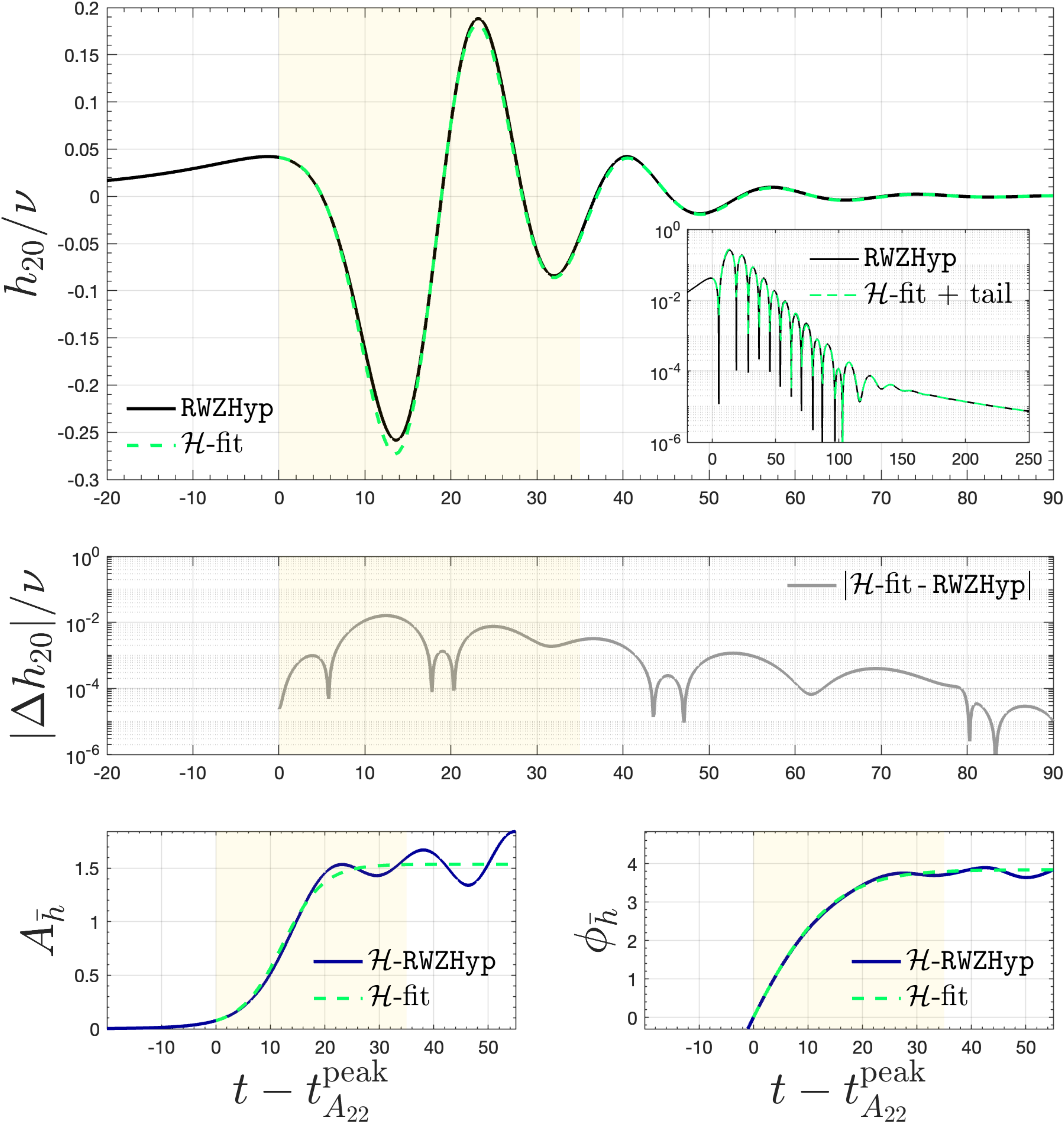}
    \caption{Upper panel: (2,0) mode for the Schwarzschild configuration. We show 
    the numerical waveform (black) and the Hilbert fit (dashed green). 
    Middle panel: corresponding residual. 
    Bottom panels: QNM-rescaled amplitudes and phase
    of the complexified numerical (2,0) mode (blue) and corresponding Hilbert fits (dashed green). 
    The vertical bands mark the ${\cal H}$-fitting interval. 
    }
  \label{fig:schw_hilbert_l2m0}
\end{figure}

\lettersection{Phenomenology}
%
\begin{figure}[t]
  \centering 
    \includegraphics[width=0.48\textwidth]{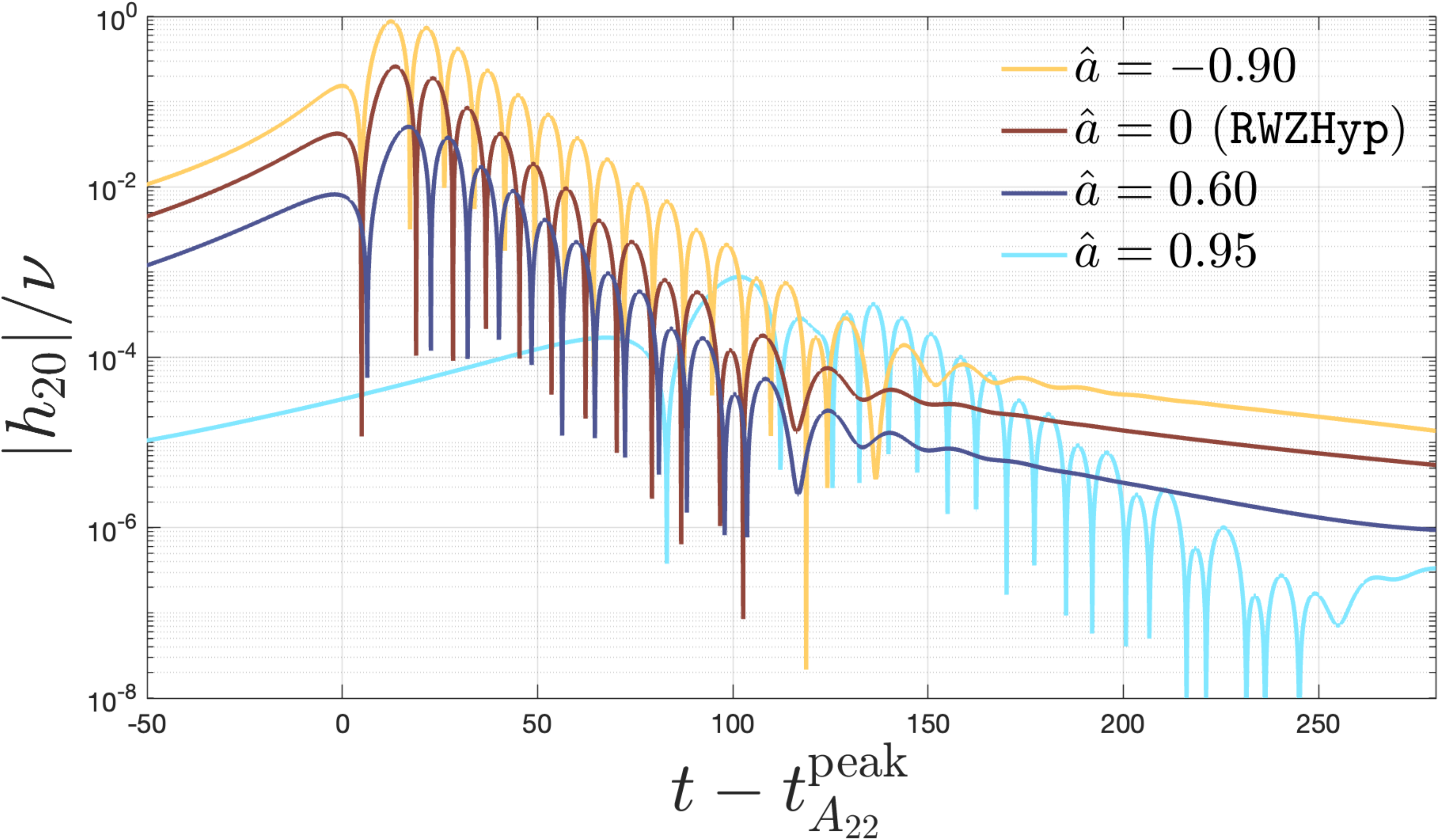}
    \caption{(2,0) modes generated by quasi-circular evolutions in Schwarzschild and Kerr for different
    spin parameters $\hat{a}$. Aligned with respect to $\tA22$. }
  \label{fig:kerr_phenom}
\end{figure}
In the test-mass limit, the relevance of the (2,0) mode changes substantially 
with the dimensionless Kerr spin $\hat{a}$, as shown in Fig.~\ref{fig:kerr_phenom}. 
Indeed, the (2,0) mode is enhanced for $\hat{a}<0$, while it is suppressed for high positive spins. 
This can be understood by noting that the $m=0$ multipoles are radial modes, and that 
the last stable orbit is larger for retrograde than for prograde orbits~\cite{Ori:2000zn},
thus leading to a more radial plunge.
Moreover, the peak of the signal is strongly delayed with respect to $\tA22$, 
as shown by the $\hat{a}=0.95$ case. 
We also observe that the tail~\cite{Price:1971fb,Leaver:1986gd,Andersson:1996cm} 
of the $m=0$ modes is more relevant than for $m>0$ ones, as expected\footnote{Indeed, the tail is enhanced by radial motion; see, \eg{}, Ref.~\cite{DeAmicis:2024not} and Eq.~(22) therein or Refs.~\cite{DeAmicis:2024eoy,Ma:2024hzq}.}.
Indeed, in Schwarzschild the transition from \ac{QNM} to tail 
for the dominant mode occurs at $t \sim t_{A_{22}}^{\rm peak}+ 230$, when $A_{22}/\nu\sim 5 \times 10^{-9}$,
while for the (2,0) mode it occurs at $t \sim t_{A_{22}}^{\rm peak}+ 120$, when $A_{20}/\nu\sim 5\times 10^{-5}$.
For the \RWZ{} simulation here considered, we find that the tail is well
described by a power-law $h_{20}^{\rm tail} \propto (t-t_{A_{22}}^{\rm peak})^{-2.787}$ 
(see inset of Fig.~\ref{fig:schw_hilbert_l2m0}).
However, in general, the tail can have rather complicated behaviors,
as extensively discussed in Ref.~\cite{DeAmicis:2024not}. 
Eccentric \RWZ{} simulations show that the tail enhancement due to the orbital eccentricity 
observed for the $m>0$ modes~\cite{Albanesi:2023bgi,Cardoso:2024jme,DeAmicis:2024not,Islam:2024vro} 
occurs also for the (2,0) mode in Schwarzschild.  

By inspecting the SXS catalog~\cite{SXS:catalog}, we observe an enhancement of the (2,0) oscillatory 
mode for binaries with anti-aligned spins also in the comparable-mass case. 
However, as the symmetric mass ratio increases, the null memory
contribution gains importance, as it is linked to the radiated energy flux. For the same reason,
the memory contribution is enhanced for spin-aligned binaries, where, instead, the oscillatory
part is suppressed. 
The accurate modeling of both contributions is thus needed. 

\lettersection{Hilbert post-merger model}
%
\ac{IMR} models have to be completed
with analytical ringdown models built employing 
\ac{NR} simulations~\cite{Buonanno:2006ui,Damour:2007xr,Baker:2008mj,Damour:2014yha}. 
In particular, Ref.~\cite{Damour:2014yha} proposed to factorize 
the fundamental \ac{QNM} from the numerical multipoles,
thus obtaining a QNM-rescaled mode. The latter is defined as 
$\bar{h}(\tau) \equiv e^{\sigma_1 \tau + i \phi_0} h(\tau) $,
where $\sigma_1 = \alpha_1 + i \omega_1$ is the fundamental
QNM, $\phi_0$ is the phase at the starting time of the fit $t_{\rm fit}^0$, $\tau = (t - t_{\rm fit}^0)/M_f$,
and $M_f$ is the mass of the remnant.
This signal thus still contains all the  source-driven transient contributions, higher QNM overtones,
and eventual non-linearities that are in numerical waveforms. 
The closed-form description is then achieved by fitting the amplitude and the phase 
of the rescaled waveform, $\bar{h} = A_{\bar{h}} e^{i \phi_{h}}$. 
The model's accuracy relies in the simple shape of the 
amplitude and phase of $\bar{h}$, which can be easily fitted with 
activation-like monotonic functions.

While this approach has been successfully adopted for complex modes, it cannot be 
readily applied to the real $m=0$ modes due to its inherently complex nature. 
Moreover, complex signals allows us to impose continuity conditions on amplitude and frequency,
which is important for constraining fit parameters 
and for applying \ac{NQC} corrections, as we will discuss in the next section.
We thus propose to i) complexify the $m=0$ modes, ii) apply a model
similar to the one discussed above, iii) get the physical signal
by solely considering the real part. 

To this end, we use the Hilbert transform~\cite{Hilbert:1912,Bracewell:2000} of a function $v(t)$, defined as 
$\HT{v}(t) \equiv 2/\pi \lim_{\epsilon \rightarrow 0} \int_{\epsilon}^{\infty} \frac{v(t-\tau) - v(t+\tau)}{2 \tau}d\tau$.
Its meaning is clarified by the relation with the Fourier transform ${\F}\left[v\right](\omega)$,

\begin{equation}
{\F}\left[\HT{v}\right](\omega) = -i\, {\rm sgn} (\omega) {\F}[v]\left(\omega\right),
\end{equation}
where ${\rm sgn}(x)$ is the sign function. Therefore, the Hilbert transform shifts 
the negative (positive) frequencies of $v(t)$ by $\frac{\pi}{2}$ ($-\frac{\pi}{2}$). 
Given a real signal $v_{\rm R}(t)$, the Hilbert transform can be used to create
an analytical signal $v_{\cal H}(t) = v_{\rm R}(t) - i \HT{v}(t)$. If we apply this complexification
to the $m=0$ modes, we find amplitudes and phases that {\it look like} the ones of the $m>0$ modes. 
We can thus generalize the phenomenological ringdown model 
of Ref.~\cite{Damour:2014yha} to real modes by simply working with their
complexifications.

The Hilbert transform has to be performed only when building the model; once the parameters of the 
ans\"atze have been found, we can just evaluate the closed-form post-merger waveform.

We start by applying this procedure to the (2,0) mode of the Schwarzschild case. 
It is known that for complex higher modes that are strongly delayed with respect to the (2,2) one,
it is convenient to apply the model of Ref.~\cite{Damour:2014yha} from 
$\tA22$ rather than from $t_{A_{\lm}}^{\rm peak}$~\cite{Cotesta:2018fcv}.
We thus perform our Hilbert fit starting from $t_{\rm fit}^0=\tA22$.
This choice facilitates the matching with the inspiral waveform discussed in the next section. 
The amplitude and phase of the complexified QNM-rescaled waveforms are modeled with ans\"atze similar
to the ones proposed in Ref.~\cite{Damour:2014yha}, as detailed in the \ac{SM}.
Note that some coefficients are constrained in terms of numerical quantities, $\lbrace A_0, \dot{A}_0, \omega_0 \rbrace$, 
by imposing continuity conditions. 
The results for the Schwarzschild case are displayed in Fig.~\ref{fig:schw_hilbert_l2m0}. 
In the bottom panels we report the amplitude and phase of the complexified numerical
waveform (dark blue), together with the primary Hilbert fits (dashed green). 
The choosen ans\"atze are able to catch the behavior of the complexified signal, 
yielding a real (2,0) mode (top panel) with analytical/numerical residual 
$\lesssim 0.02$ (middle panel). 
However, it should be noted that the phase template is more accurate than the amplitude one,
meaning that future works could explore different choices for the amplitude ansatz to further improve
the accuracy of this procedure. 
Moreover, the oscillations in the complexified signal observed for $t\gtrsim\tA22 + 20$ 
are likely to be, at least in part, an artefact of the transform. Indeed, they also appear when complexifying 
the real part of the (2,2) mode, while they are less evident in the corresponding physical waveform. 

The post-merger Hilbert fit can readily be applied to higher modes, Kerr
configurations, and NR simulations. While some primary fits performed over SXS simulations
are shown in the \ac{SM}, we build our model for the oscillatory contribution
using waveforms generated with \Surr{}~\cite{Varma:2018mmi}.
Once the coefficients have been found for enough configurations, we can
fit them over the parameter space. 
For the mass and spin of the remnant,
we use the fits of Refs.~\cite{Jimenez-Forteza:2016oae,Nagar:2020pcj},
while for the coefficients of the new model and $\lbrace A_0, \dot{A}_0, \omega_0 \rbrace$ we perform a hierarchical
fit. We first perform quadratic fits of test-mass and non-spinning configurations, separately.
We then complete our model by merging these two results over the parameter space identified by $(\nu,\ta)$,
where $\ta=\left(m_1\chi_1+m_2\chi_2\right)/M$ generalizes the Kerr spin. 
The explicit ans\"atze and the coefficients found are given in the \ac{SM}. 

\lettersection{Matching with inspiral waveform}
%
\begin{figure*}[t]
  \centering 
    \includegraphics[width=0.315\textwidth]{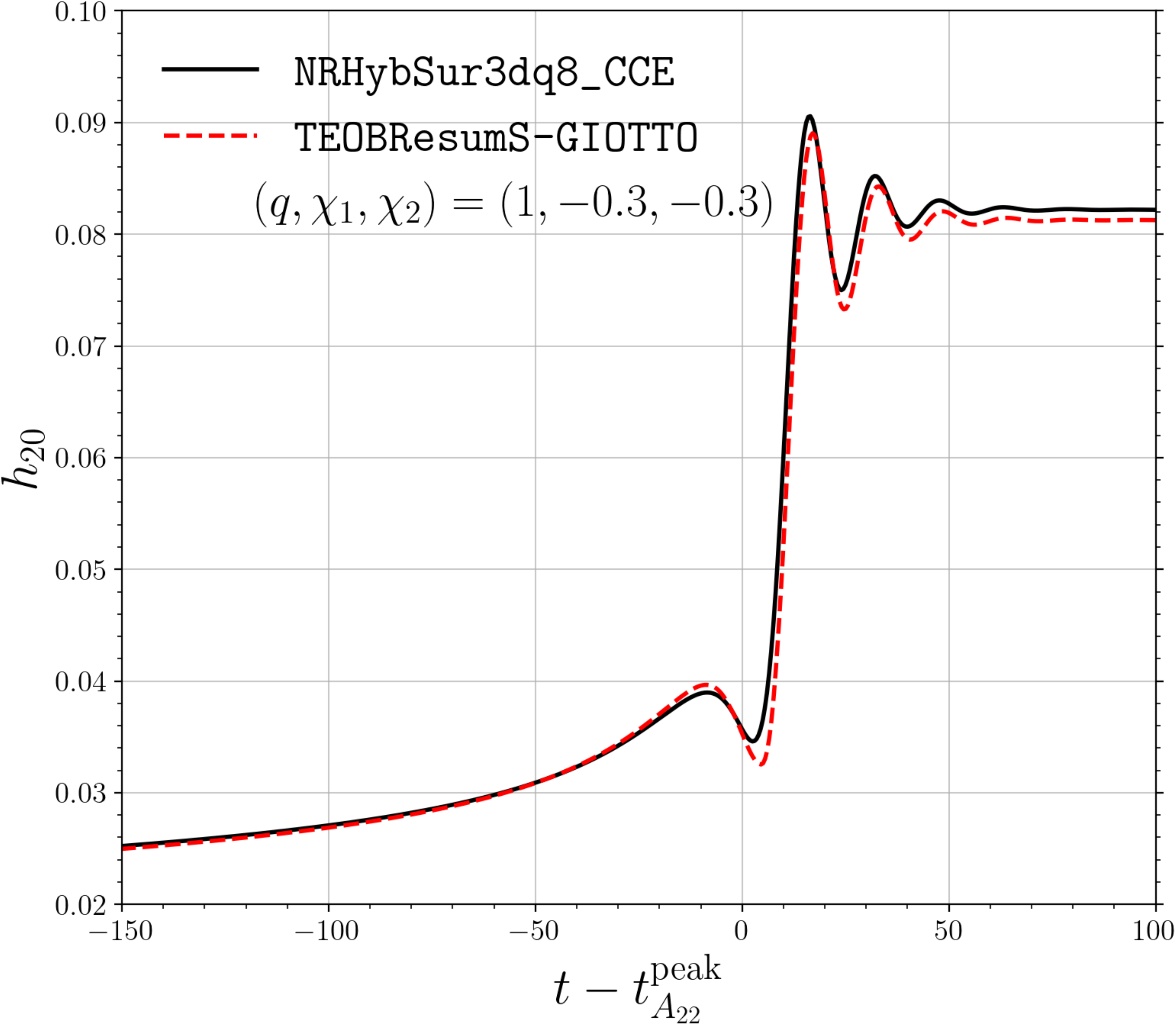}
    \includegraphics[width=0.315\textwidth]{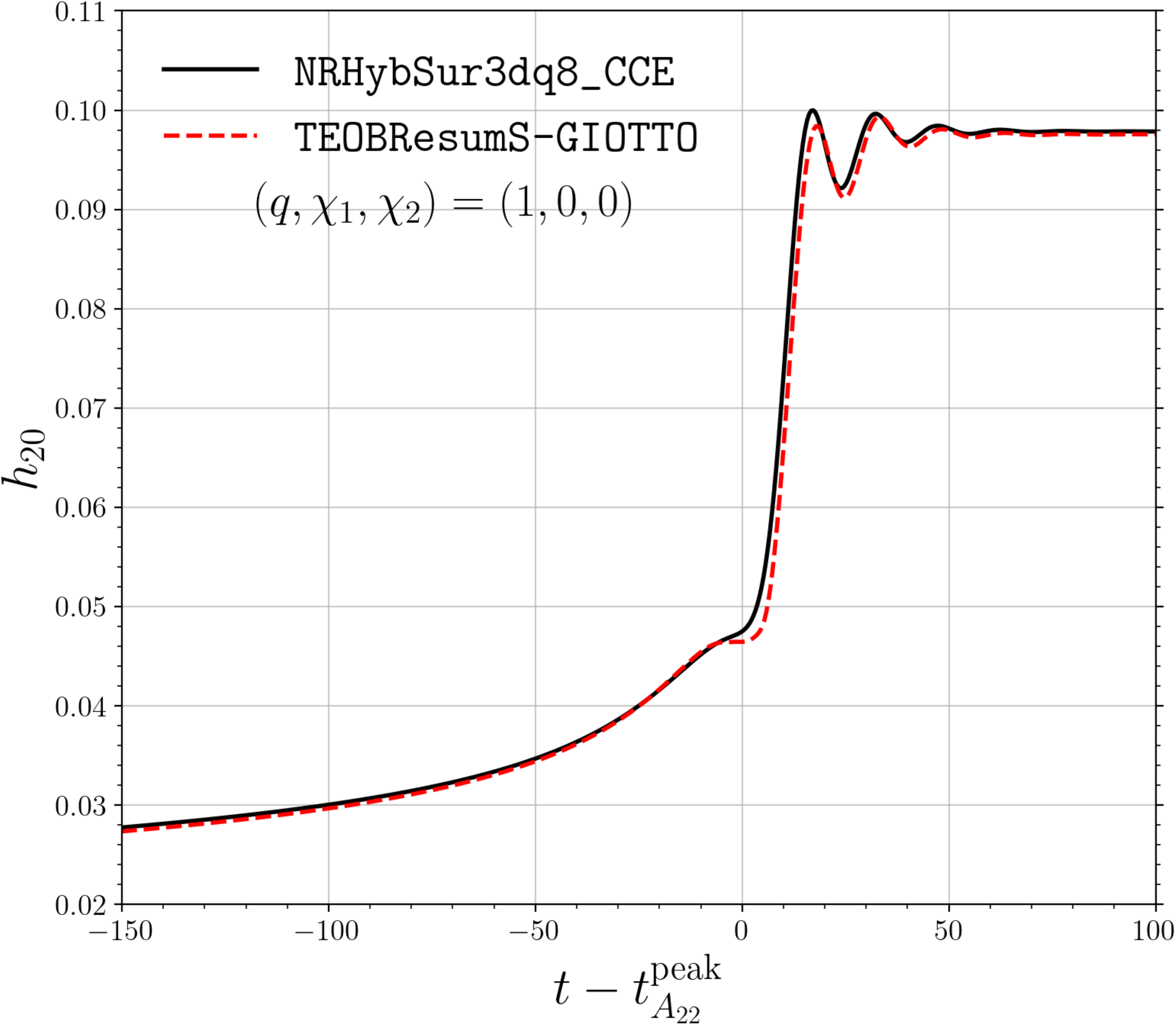}
    \includegraphics[width=0.33\textwidth]{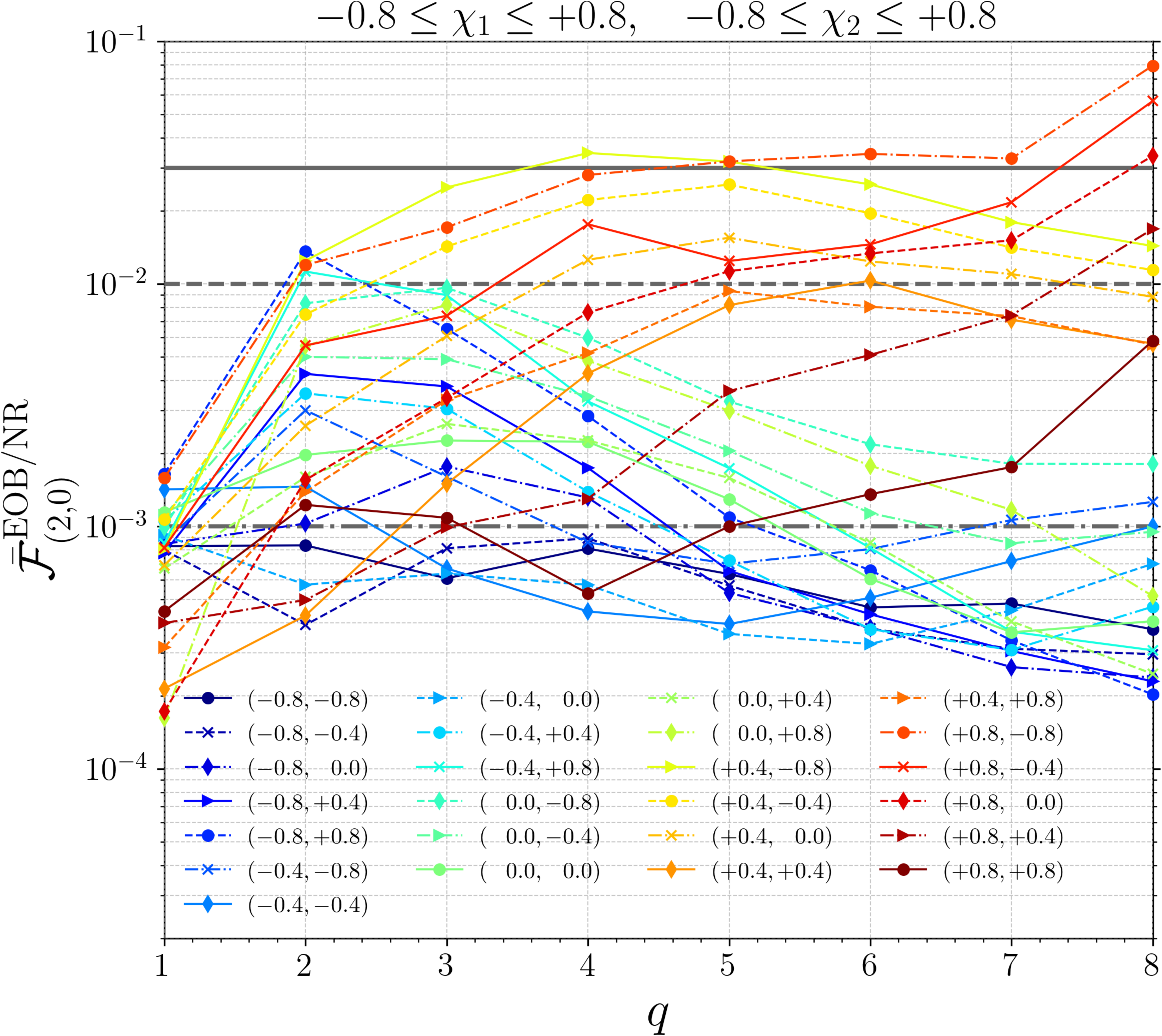}
    \caption{Comparisons of (2,0) modes given by \SurrCCE{} and the ones computed in this work using \GIOTTO{} as a baseline.
    Left and middle panels: time-domain comparisons. Right panel: LIGO-weighted mismatches of the (2,0) 
    modes for different mass ratios and spin configurations. The legend in the right panel reports $(\chi_1,\chi_2)$.}
  \label{fig:eob_vs_cce}
\end{figure*}
A formally Newtonian expression for the inspiral $m=0$ modes can be obtained following 
Ref.~\cite{Chiaramello:2020ehz}. For the (2,0) mode, this reads 
\begin{equation}
\label{eq:h20N}
h_{20}^{\rm inspl} = 4 \sqrt{\frac{2 \pi}{15}} \nu \left(r \ddot{r} + \dot{r}^2\right).
\end{equation}
Note that 2.5PN corrections for $h_{20}^{\rm inspl}$ have been recently computed~\cite{Placidi:2023ofj,Grilli:2024lfh}
and might be useful for generic orbits. However, since here we are focusing on the
plunge of quasi-circular evolutions, we do not consider them\footnote{Moreover, PN results are
known to be unreliable in the late stages of the evolution.}. 
To match this waveform to the post-merger model previously discussed, we can also complexify
$h_{20}^{\rm inspl}$ with a Hilbert transform. The matching can be then performed 
with standard EOB techniques, which involve \ac{NQC} corrections. 
Note that for NQC corrections to work effectively, well-defined and preferably monotonic amplitude and phase 
are needed. We achieve these requirements through the aforementioned complexification. 
Although this matching is a standard procedure in EOB models, we briefly discuss it 
in the \ac{SM} for completeness, along with minor technical details.

\lettersection{Null memory effects}
%
Null memory effects~\cite{Christodoulou:1991cr,Thorne:1992sdb,Blanchet:1992br,Mitman:2024uss},
which are particularly relevant for comparable mass binaries, 
can be computed using \ac{BMS} balance laws~\cite{Mitman:2020bjf}. These relations
connect the strain to the Bondi aspects
(\ie{}, the oscillatory parts discussed so far) and the radiated fluxes.
More specifically, the displacement memory in the (2,0) mode arises 
from the supertranslation conservation law, and is thus sourced
by the radiated energy flux. It can be written in terms of the most relevant multipoles as~\cite{Rossello-Sastre:2024zlr}
\begin{align}
h_{20}^{\rm memo}(t) = 
&  \frac{1}{7}\sqrt{\frac{5}{6\pi}}   \int_{t_0}^t |\dot{h}_{22}|^2 dt 
 - \frac{1}{14}\sqrt{\frac{5}{6\pi}}  \int_{t_0}^t |\dot{h}_{21}|^2 dt \cr
&+ \frac{5}{2\sqrt{42\pi}}            \int_{t_0}^t \left( \dot{h}_{22}^{\rm Re} \dot{h}_{32}^{\rm Re} + \dot{h}_{22}^{\rm Im} \dot{h}_{32}^{\rm Im} \right) \cr
&- \frac{2}{11}\sqrt{\frac{2}{15 \pi}}\int_{t_0}^t |\dot{h}_{44}|^2 dt. \label{eq:memo20}
\end{align}
Formally, one should consider $t_0\rightarrow -\infty$, but in practical application $t_0$ is the starting
time of the evolution. This means that, if we extrapolate back in time, we get the unphysical property 
$h_{20}^{\rm memo} \neq 0$ for $t\rightarrow -\infty$. This can easily fixed by matching this memory
contribution to the PN formula~\cite{Favata:2008yd,Cunningham:2024dog}, 
which correctly vanishes at zero frequency. Note that Ref.~\cite{Rossello-Sastre:2024zlr} 
followed a similar approach. Technical details on this matching are discussed in the \ac{SM}.
Therefore, given the $m>0$ EOB multipoles, we can compute the full $(2,0)$ mode as
$h_{20} = h_{20}^{\rm osc} + h_{20}^{\rm memo}$, where $h_{20}^{\rm osc}$ 
is the oscillatory contribution previously discussed.

\lettersection{EOB/NR comparisons}
We now apply the methodology here introduced to the EOB model \GIOTTO{}~\cite{Nagar:2023zxh} 
and compare it with \ac{NR} results. 
To avoid ambiguities linked to different BMS frames, we consider
the surrogate \SurrCCE{}~\cite{Yoo:2023spi}, whose waveforms 
are already in the PN frame, as the EOB ones. 
Time-domain comparisons for two meaningful equal mass cases are reported in Fig.~\ref{fig:eob_vs_cce},
showing that our model is able to capture the main features of the 
numerical waveform.
Since the ringdown modeling of the higher modes enter in the computation of the final EOB offset,
improving their description would yield an offset closer to the numerical one.
This is however beyond the scope of this work, and thus deferred to future investigations. 
Some additional time-domain comparisons are reported in the \ac{SM}. 
To assess the accuracy more systematically, we also consider frequency-domain mismatches for the (2,0) mode
for different mass-ratios and spins in the right panel of Fig.~\ref{fig:eob_vs_cce}. 
We consider the noise curve of Advance LIGO~\cite{aLIGODesign_PSD} and a reference mass of
$50 M_\odot$. More technical details are reported in the \ac{SM}.
The model is generally accurate,
especially for equal-mass binaries, where all the mismatches are below $2\times 10^{-3}$, 
and for high mass-ratio cases with low spins. However, higher mismatches are obtained for 
configurations where $\chi_1$ and $\chi_2$ strongly differ. 
This can be understood by considering that we reduce the dimensionality of the fits across the parameter space by 
employing the effective spin $\tilde{a}_0$ rather than the two individual spins $\chi_{1,2}$ separately.

We conclude by discussing the quadrupolar
strains anticipated in Fig.~\ref{fig:eobnr_strain}, which are computed for an observer 
with line of sight that maximizes the contribution of the (2,0) mode 
($\Theta=\frac{\pi}{2}$). We compare our EOB model
with the results from \SurrCCE{} and the equal mass nonspinning simulation
of the \texttt{ExtCCE} SXS catalog~\cite{Mitman:2021xkq}.
Note that the SXS waveform is not in the PN frame, but we can obtain a good agreement 
with the surrogate by i) mapping it to the superrest frame with the package 
\texttt{scri}~\cite{Boyle:2013nka,Boyle:2014ioa,Boyle:2015nqa,mike_boyle_2020_4041972}, ii) shifting
its (2,0) mode in order to match the final offset of the surrogate model. 
A more accurate, though more complex, procedure would involve mapping the 
SXS waveform to the PN frame as detailed in Refs.~\cite{Mitman:2021xkq,Mitman:2022kwt}.
We remark that no additional offset-shifting is performed for the EOB and \SurrCCE{} waveforms. 
The agreement between the three shows the goodness of \GIOTTO{} completed with the (2,0) mode as discussed in this work.
The EOB/NR mismatch for the complete quadrupolar waveform in this case is $\bar{\cal F} = 6\times 10^{-4}$.


\lettersection{Conclusions}
We have proposed a method to include the description of the real $m=0$ modes 
in the EOB framework, thus completing the multipolar structure of
\ac{IMR} EOB waveforms for spin-aligned quasi-circular \acp{BBH}. 
In particular, we have discussed how to model
the oscillatory post-merger of such modes by first complexifying them with
a Hilbert transform, and then using a post-merger ansatz based on
Ref.~\cite{Damour:2014yha}, as typically done for $m>0$ modes. 
We thus effectively include source-driven 
and higher QNM overtones contributions. 
This model has been applied to a set of test-mass simulations, hence
showing its applicability in the linear perturbation regime. We then focused on the (2,0) modes and 
discussed how to employ \ac{NQC} corrections to match the Hilbert post-merger model 
to the inspiral EOB waveform obtained with the approach introduced in Ref.~\cite{Chiaramello:2020ehz}. 
The close-form description of the (2,0) post-merger for the comparable mass 
case has been achieved by performing hierarchical fits for different mass ratios and effective spins
using waveforms from \Surr{}~\cite{Varma:2018mmi}. These fits correctly include the test-mass limit.
We have then discussed how to include null memory effects in EOB models, thus providing a 
complete description of the (2,0) mode. Following this prescription, we have complemented \GIOTTO{} and
tested its validity against the surrogate \SurrCCE{}~\cite{Yoo:2023spi}. We also 
performed a comparison with the equal mass nonspinning configuration of the SXS catalog 
of Ref.~\cite{Mitman:2021xkq}, showing a remarkable agreement.

Future works will focus on refining the methodology here introduced and 
on employing it to complete EOB models where the relevance of the (2,0) mode
is even more enhanced than for spin-aligned quasi-circular binaries, 
such as precessing and eccentric models~\cite{Gamba:2024cvy}. 

\lettersection{Acknowledgment}
S.A. is grateful to R. Gamba, S. Bernuzzi and A. Nagar for useful discussions through the 
whole development of this work. S.A. also thanks G. Carullo and M. De Amicis
for useful comments and suggestions on the draft. 
S.A.~acknowledges support from the Deutsche Forschungsgemeinschaft (DFG) project ``GROOVHY'' (BE 6301/5-1 Projektnummer: 523180871).
The EOB model \GIOTTO{} and its version for generic orbital motion, \DALI{}, are publicly available at 
{\footnotesize \url{https://bitbucket.org/teobresums/teobresums/src/}}. 
However, the procedure for computing the (2,0) mode described in this work has not yet been implemented in the public code.
In Fig.~\ref{fig:kerr_phenom}, we used a scientific colormap provided by Ref.~\cite{crameri_2023_8409685}.


\bibliography{refs20250627, refs_loc20250627}



\renewcommand{\thesubsection}{{S.\arabic{subsection}}}
\setcounter{section}{0}

\section*{Supplemental material}

\lettersection{Primary ans\"atze for the Hilbert fit} 
%
\begin{figure*}[t]
  \centering 
    \includegraphics[width=0.32\textwidth]{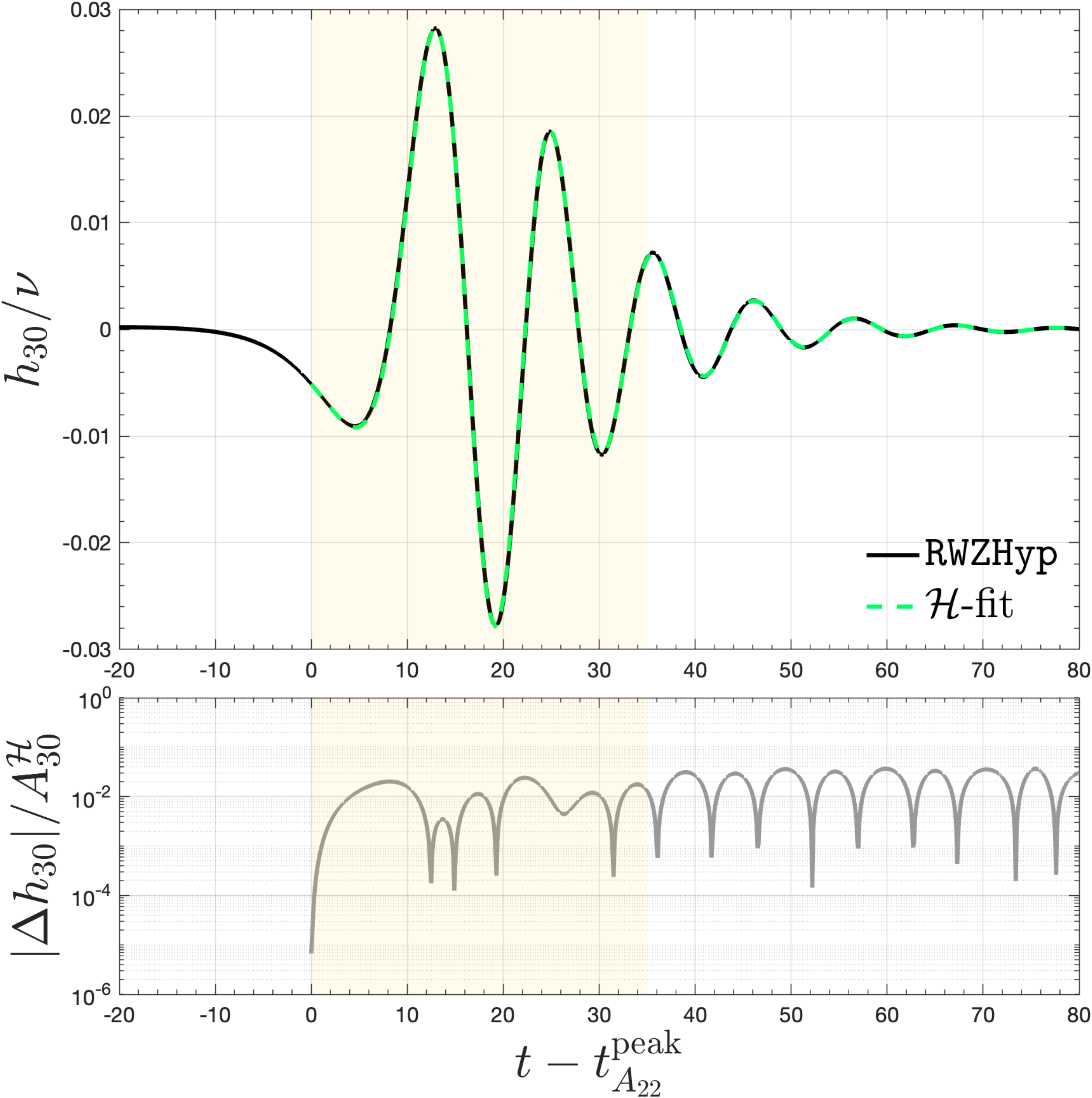}
    \includegraphics[width=0.32\textwidth]{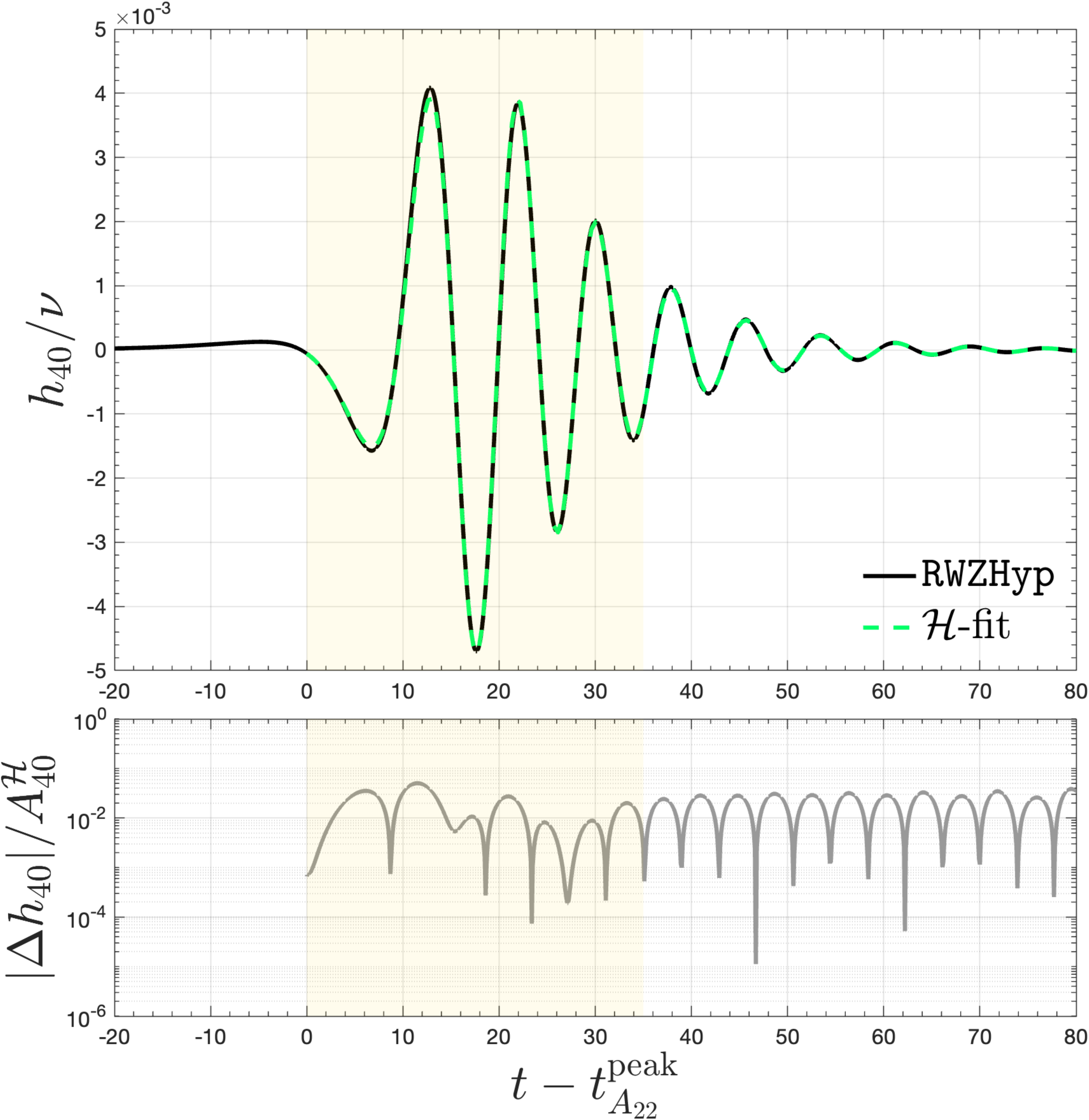}
    \includegraphics[width=0.32\textwidth]{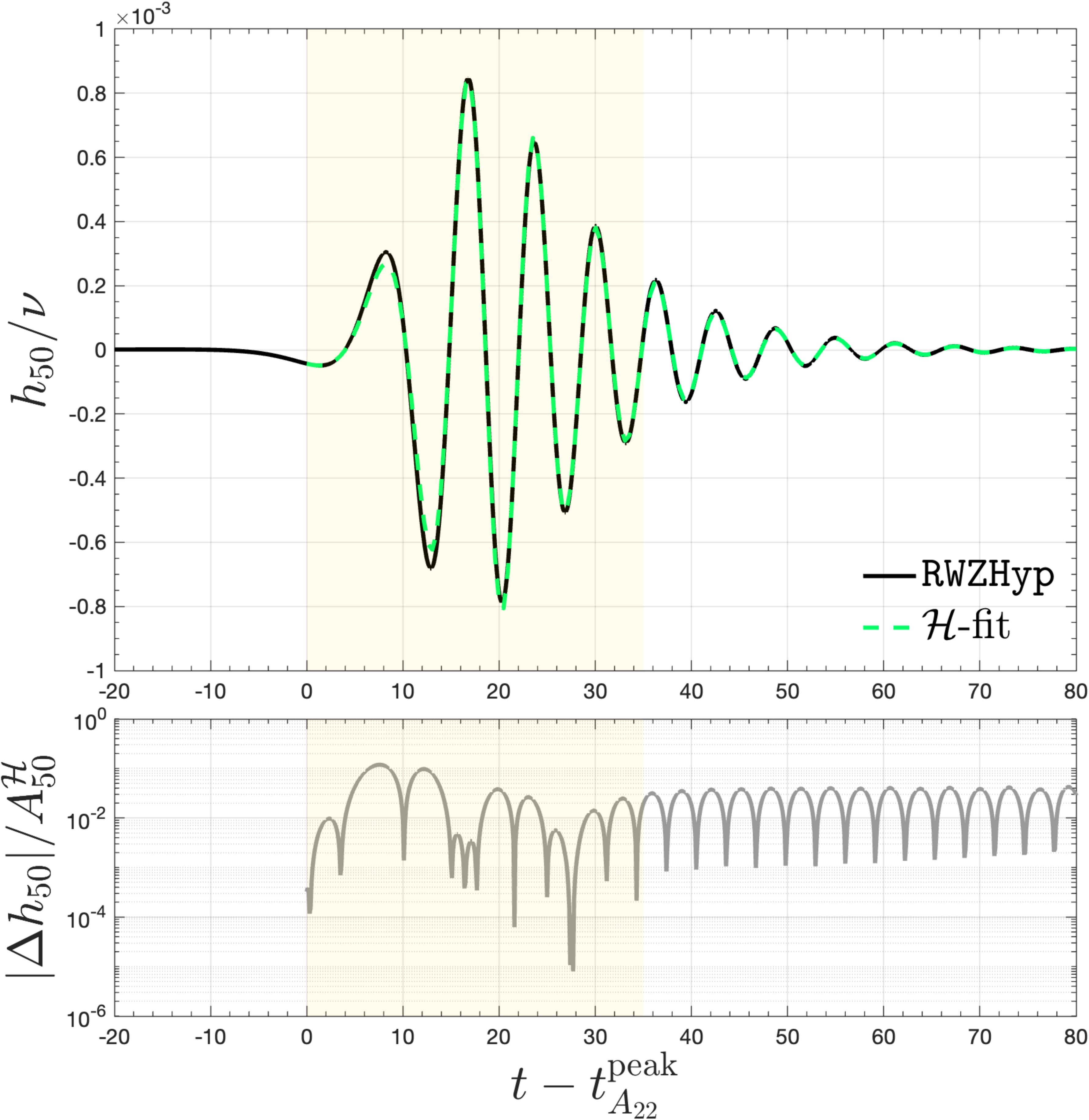}
    \caption{\RWZ{} (3,0), (4,0) and (5,0) waveforms generated by the
    inspiral-plunge-merger of a test particle in Schwarzschild (black), together
    with the corresponding Hilbert fits (dashed green). In the bottom panels we report
    the analytical/numerical residuals normalized with the amplitude of the complexified 
    signals. The vertical bands mark the ${\cal H}$-fitting interval.
    }
  \label{fig:schw_hilbert_HMs}
\end{figure*}
We report here the explicit expressions used for the primary fits 
of the amplitude and phase of the QNM-rescaled waveform~\cite{Damour:2014yha,Cotesta:2018fcv}. They read
\begin{align}
\label{eq:templateA}
A_{\bar{h}}    & =  c_1^A    \tanh\left( c_2^A \tau + c_3^A \right) + c_4^A, \\
\label{eq:templatePhi}
\phi_{\bar{h}} & = -c_1^\phi \ln  \left( \frac{1+c_3^\phi e^{-c_2^\phi \tau} }{1 + c_3^\phi} \right),
\end{align}
where $\tau=(t-t_{\rm fit}^0)/M_f$, and $M_f$ is the mass of the remnant. By imposing continuity conditions at $t_{\rm fit}^0$, we constrain the following
parameters:
\begin{align}
c_1^A    &= \left(\dot{A}_0  + \alpha_1 A_0\right) \frac{1}{c_2^A} \cosh( c_3^A )^2, \\
c_4^A    &=  A_0   - \left(\dot{A}_0 + \alpha_1 A_0\right)\frac{1}{c_2^A} \cosh(c_3^A) \sinh(c_3^A), \\
c_1^\phi &=  \frac{1+c_3^\phi}{c_2^\phi c_3^\phi} \left( \omega_1 - M_f \omega_0 \right),
\end{align}
where we recall that the fundamental QNM frequency is denoted as $\sigma_1 = \alpha_1 + i \omega_1$,
while $\omega_0$ is the waveform frequency at $t=t_{\rm fit}^0$. The 
(2,0) multipole has been already discussed in the main text. In Fig.~\ref{fig:schw_hilbert_HMs}
we report, instead, the results for the (3,0), (4,0), and (5,0) modes in the Schwarzschild case. We also show the
residuals in the bottom panels, but since the amplitude
of the modes varies of  different order of magnitudes, we normalize them with the amplitude of the 
complexified signal. As can be seen, the accuracy of the Hilbert fit remains
consistent also for the higher modes. For Kerr, we consider primary fits of the (2,0) mode up to 
$\hat{a}=0.6$. The coefficients found are reported in Table~\ref{tab:sims_testmass}.

\begin{figure*}[t]
  \centering 
    \includegraphics[width=0.32\textwidth]{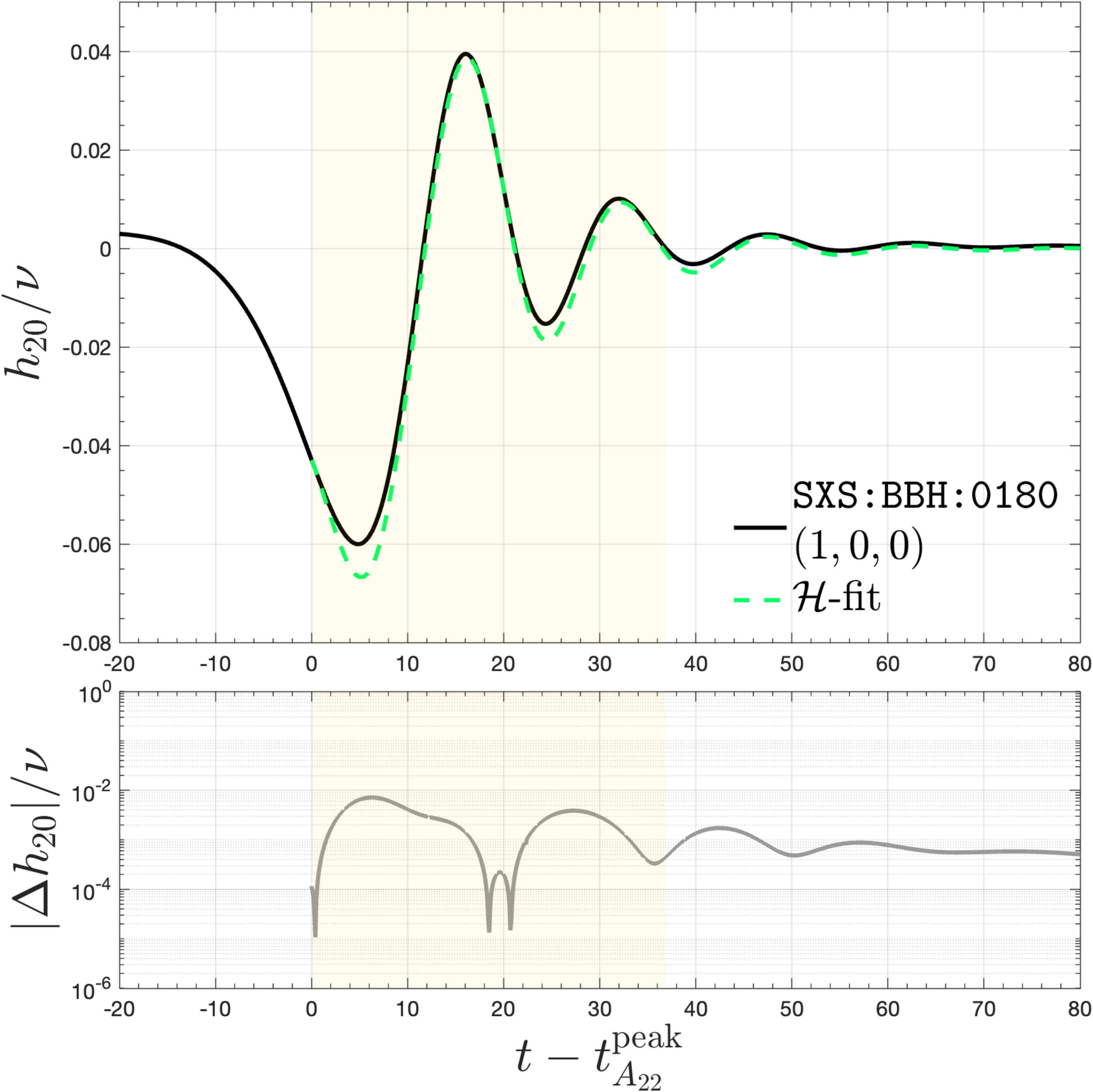}
    \includegraphics[width=0.32\textwidth]{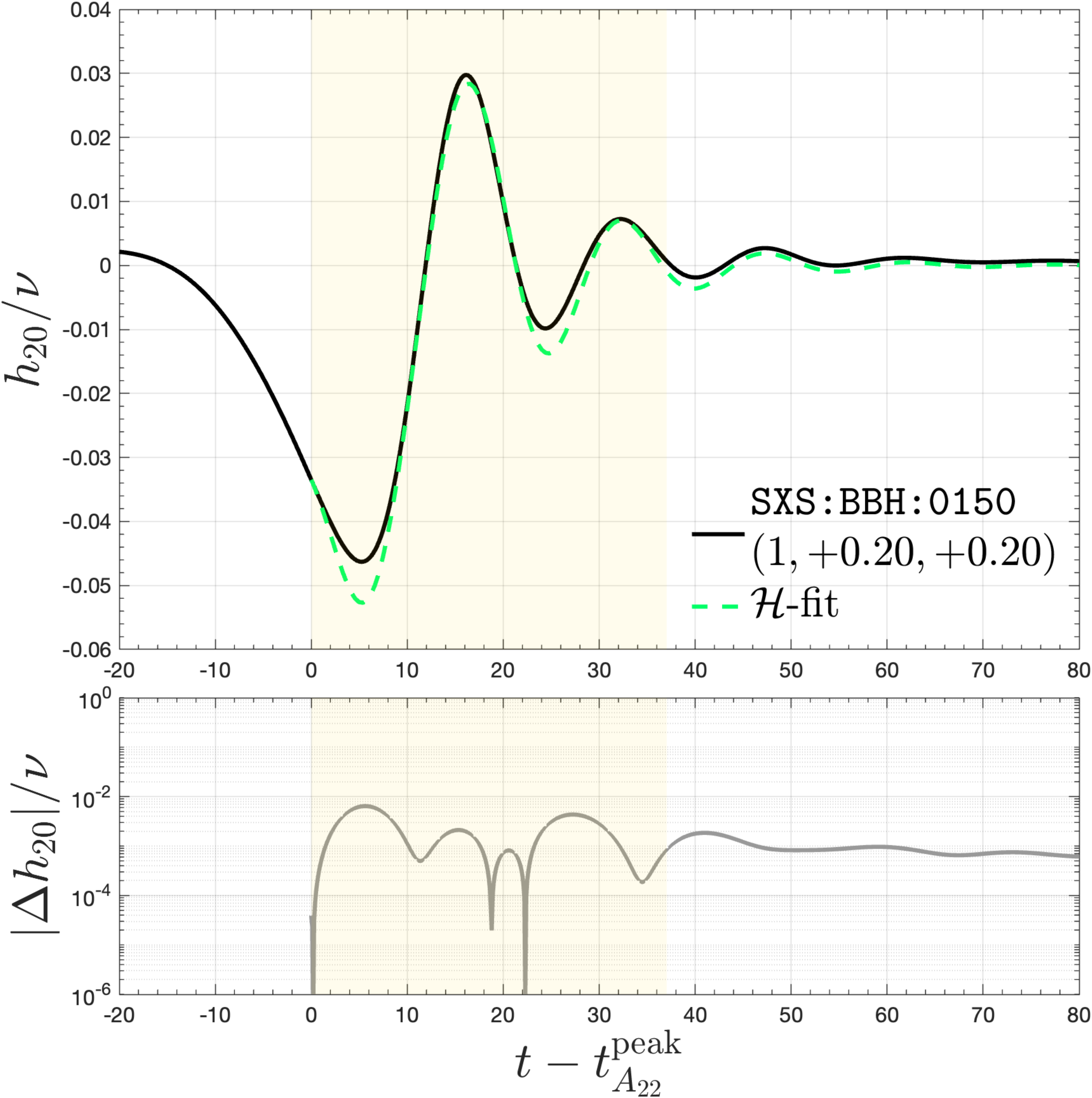}
    \includegraphics[width=0.32\textwidth]{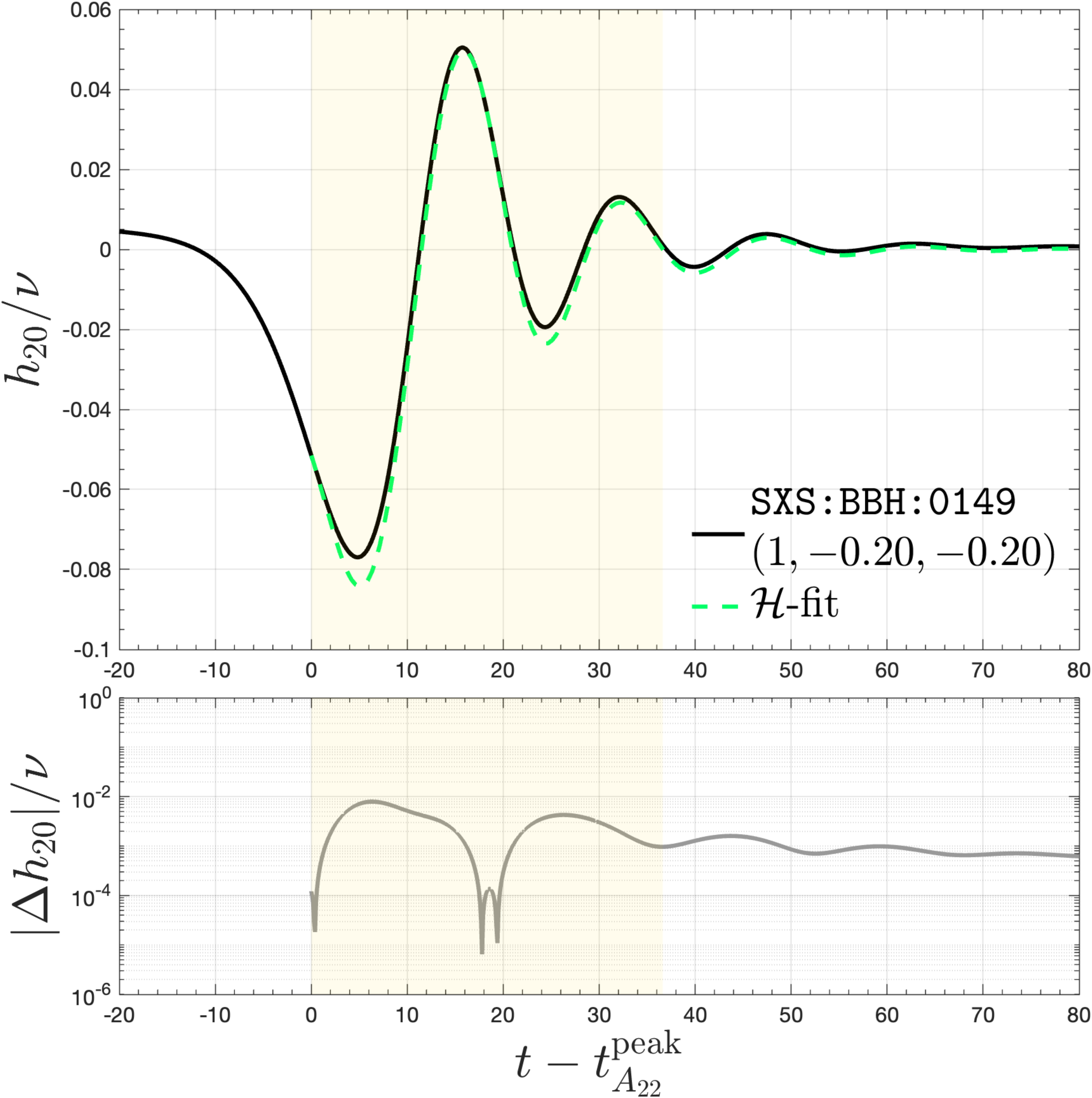}
    \includegraphics[width=0.32\textwidth]{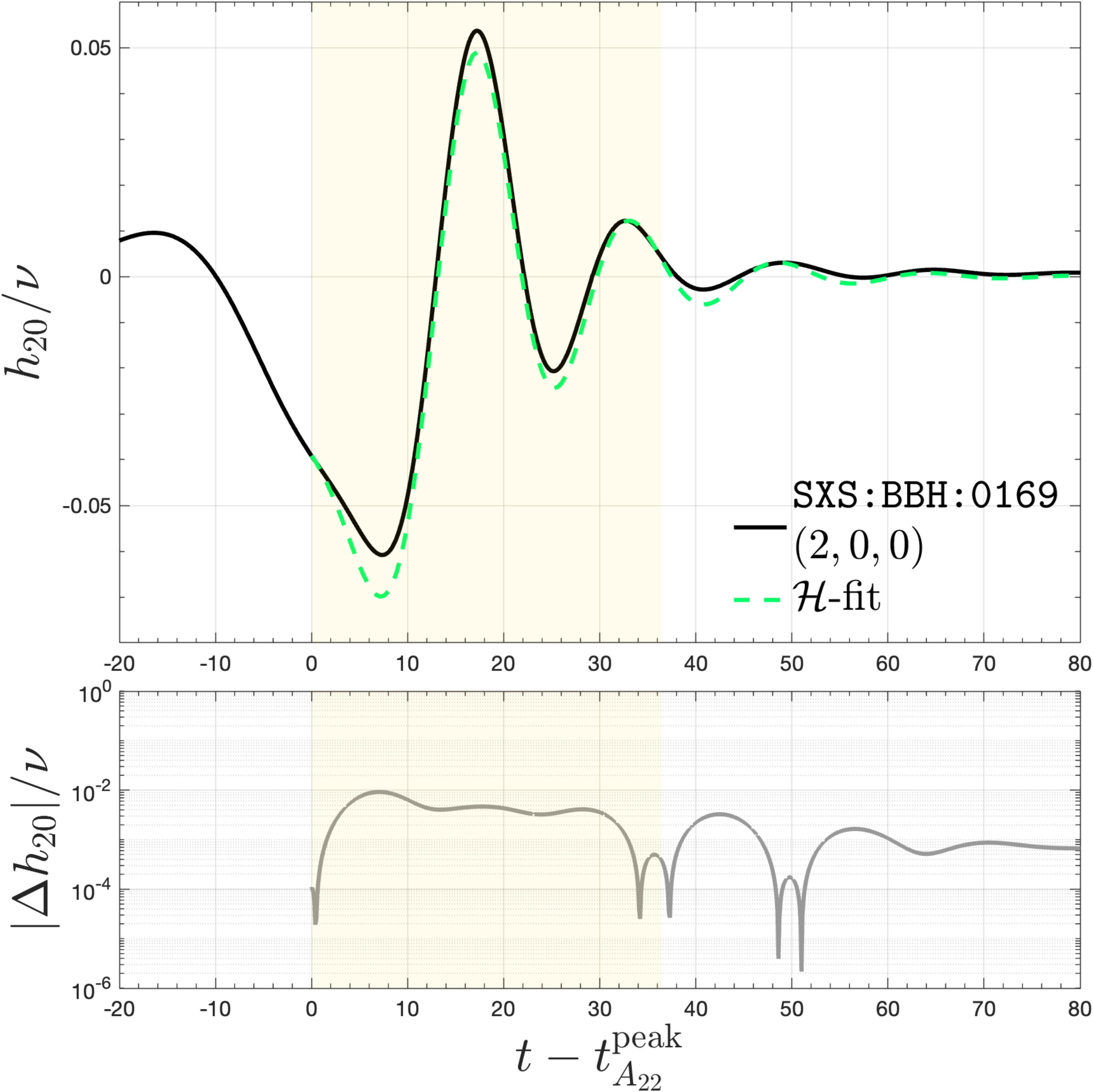}
    \includegraphics[width=0.32\textwidth]{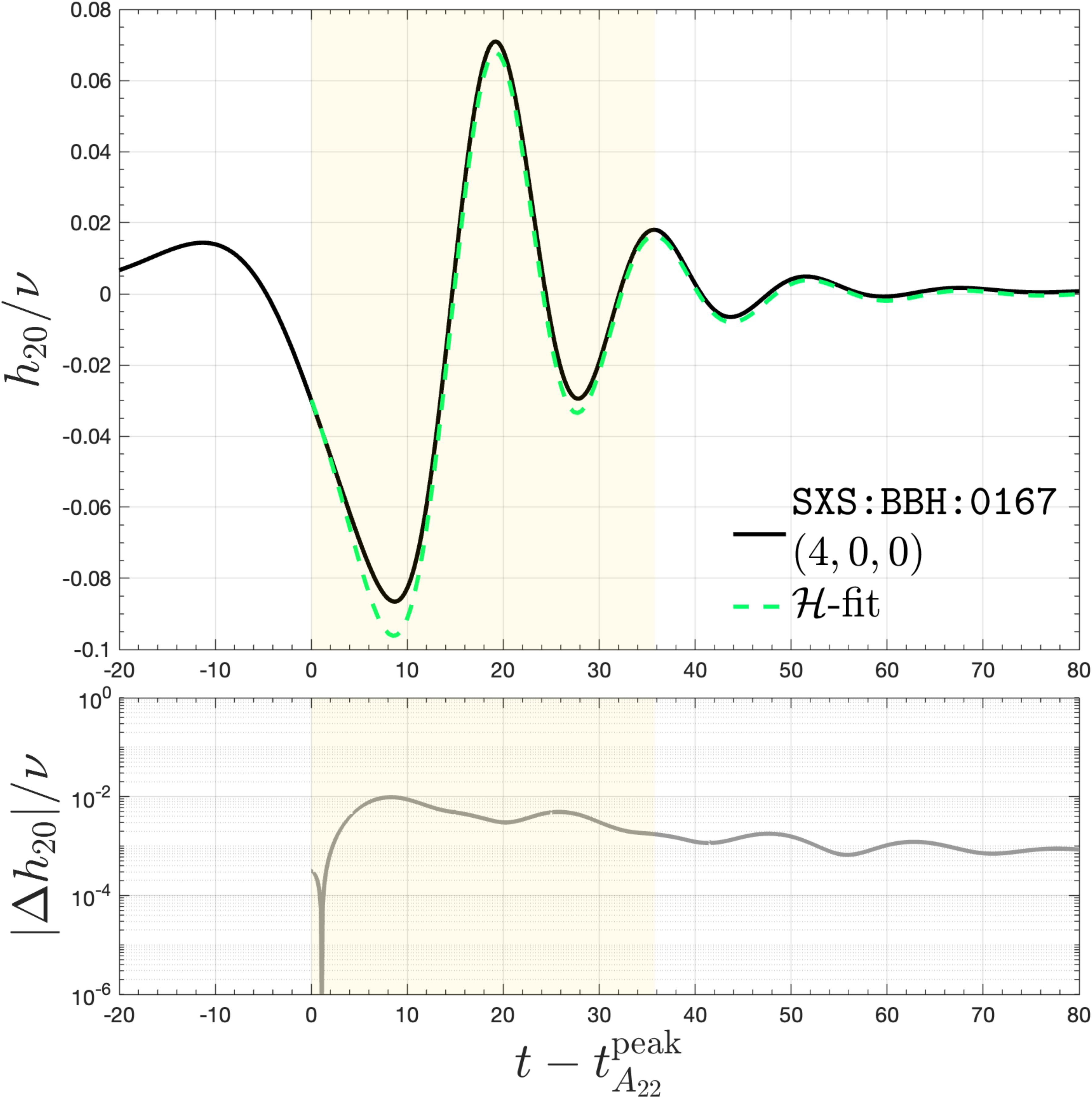}
    \includegraphics[width=0.32\textwidth]{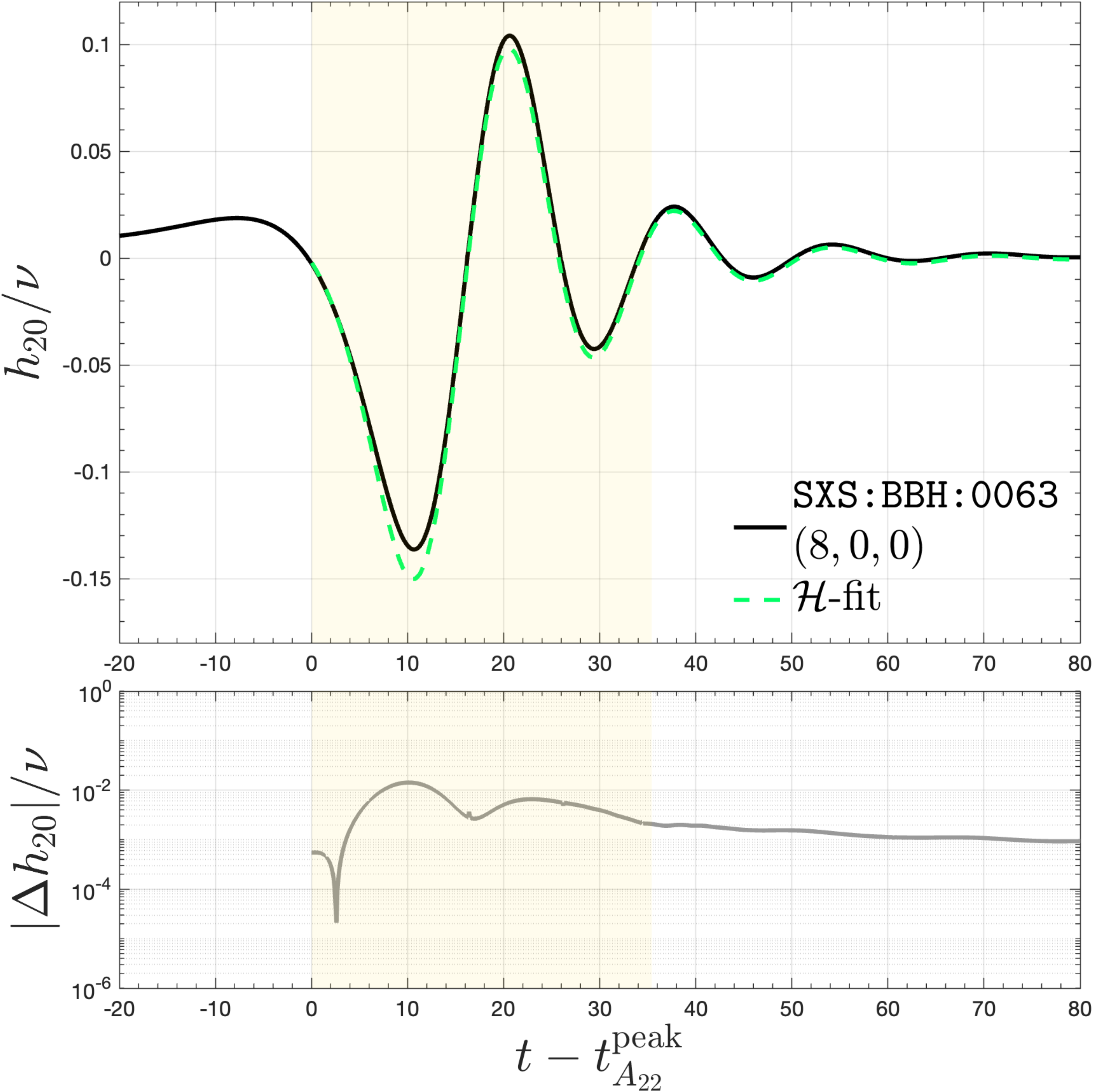}
    \includegraphics[width=0.32\textwidth]{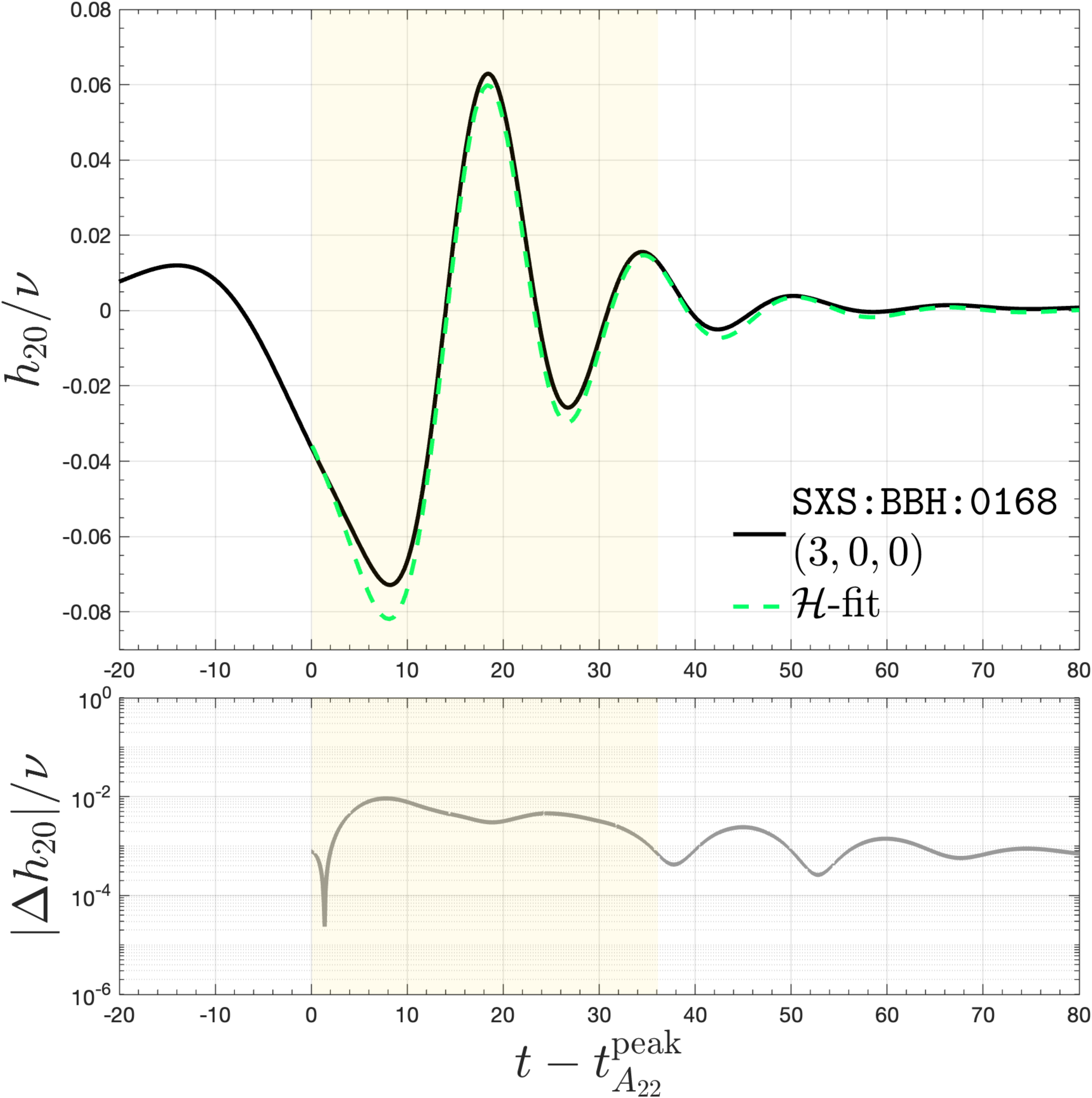}
    \includegraphics[width=0.32\textwidth]{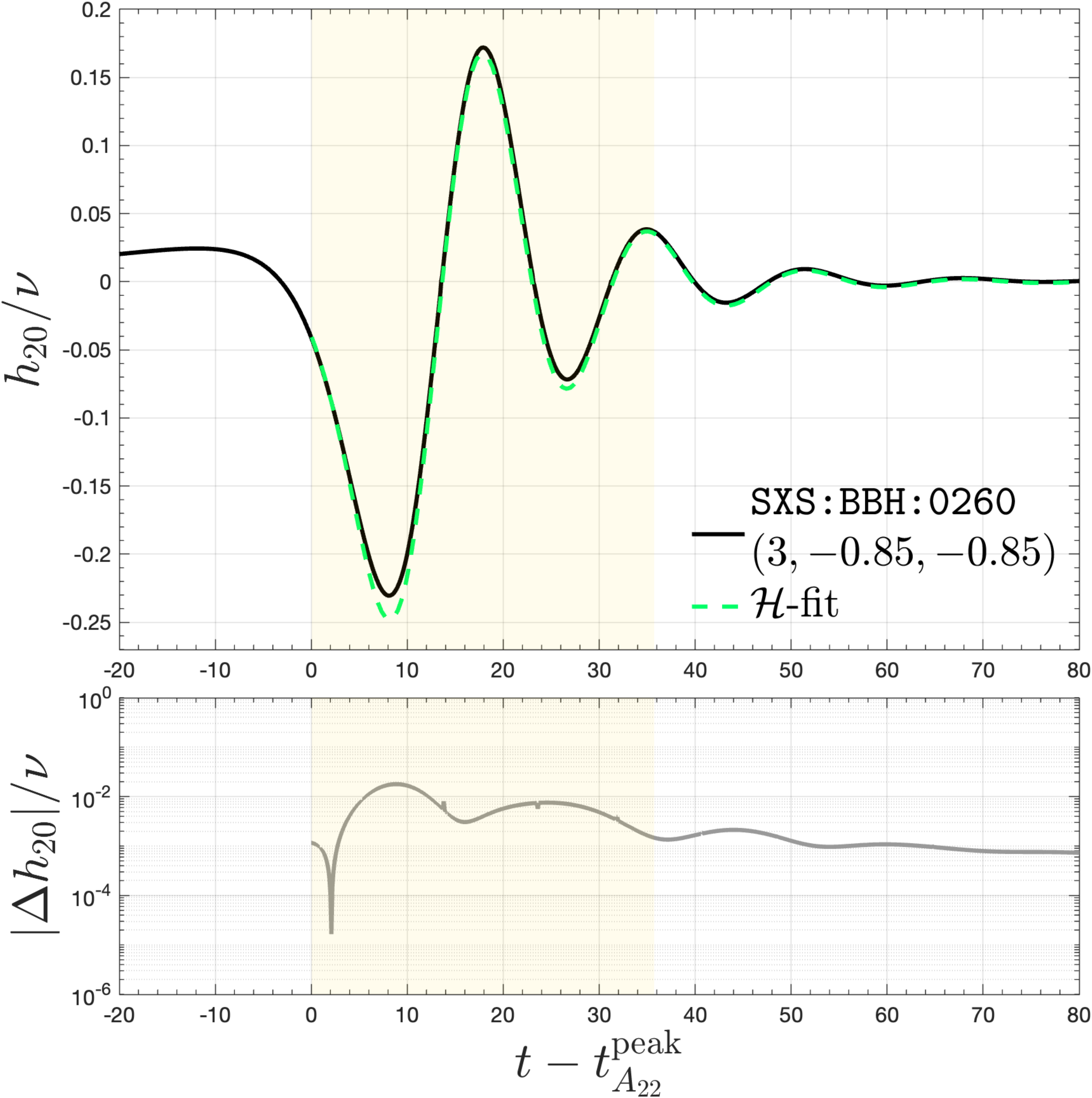}
    \includegraphics[width=0.32\textwidth]{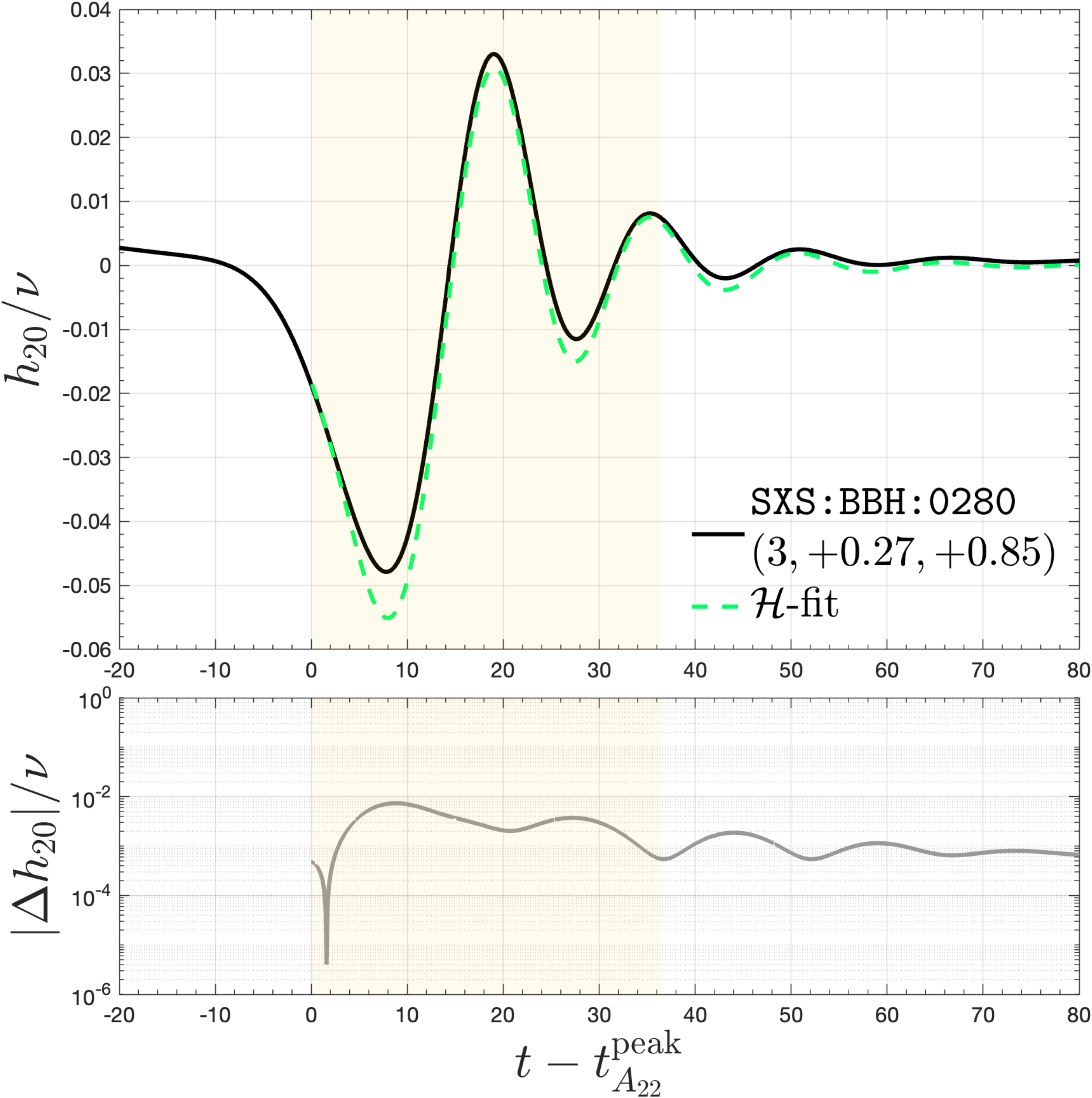}
    \caption{Primary Hilbert fits of the (2,0) mode (dashed green) performed for different SXS configurations (black).
    Residuals in the bottom panels. The vertical bands mark the ${\cal H}$-fitting intervals. Note the different vertical scales.}
  \label{fig:hilbert_fit_sxs}
\end{figure*}
We now turn our attention to the comparable mass case. 
We report some significant cases of Hilbert fits performed over SXS data of 
the public catalog~\cite{SXS:catalog} in Fig.~\ref{fig:hilbert_fit_sxs}. 
The numerical waveforms are extracted at finite distance and then extrapolated, so that null
memory effects are not captured. In other words, these numerical data 
only represent the Bondi mass aspect and, as a consequence, can be fitted with a purely oscillatory ansatz. 
The residuals between the numerical modes and the Hilbert fits are reported in the bottom panels (gray). 
While the phase of the oscillations is well reproduced, there is a larger discrepancy for the amplitude.
A similar, but less evident, discrepancy is observed also in the Schwarzschild case,
as already discussed in the main text (cfr. with Fig.~\ref{fig:schw_hilbert_l2m0}).
This is a further indication that future works could try different closed-form
representations for the QNM-rescaled amplitude $A_{\bar{h}}$. 
However, it should be also noted
that the \ac{NR} modes do not approach zero exponentially in the late ringdown, thus indicating that improvements might be needed
on the numerical side too\footnote{This slower decay could be linked to a 
physical power-law tail, but it may be also related to the residual drift of the center of mass
observed in SXS simulations~\cite{Woodford:2019tlo}.}. 
Moreover, note that the nonspinning configurations
with $q=2,3$ have unexpected behaviors before $\tA22$. 
Given these considerations, we decide to employ waveforms generated by the surrogate \Surr{}~\cite{Varma:2018mmi}, 
rather than using directly SXS waveforms. 
Therefore, for the comparable mass configurations, we select systems 
with $q\in\left[1,8\right]$ and spins $\chi \in [-0.8,0.8]$; for each $(q,\chi)$-pair,
we consider $(\chi_1,\chi_2)=\{ (\chi,0), (\chi,\chi), (0,\chi) \}$. 
Using the surrogate thus facilitates the location of waveforms
in the parameter space, improving the quality of the hierarchical global fits.
Alternative strategies to be investigated in the future 
include the computation of the Bondi mass aspect from CCE data by removing the displacement
memory contribution using Eq.~\eqref{eq:memo20}, mapping the SXS data to the superrest frame, 
or considering also data from different catalogs,
such as the RIT one~\cite{Healy:2017psd,Healy:2019jyf,Healy:2020vre,Healy:2022wdn}.

Finally, we also explored the amplitude ansatz introduced in 
Ref.~\cite{Albanesi:2021rby}, which reads 
\begin{equation}
\label{eq:templateA_c5}
A_{\bar{h}} = \left( \frac{c_1^A}{1+e^{-c_2^A \tau + c_3^A}} + c_4^A   \right)^{ \frac{1}{c_5^A} }.
\end{equation}
We tested different combinations of constrained parameters, both constraining only
two parameters using $A_0$ and $\dot{A}_0$, and three parameters using $\ddot{A}_0$ too. While 
the fitted amplitude was more accurate for the test-mass primary fits, the global fits over 
the parameter space were not stable, in the sense that the argument of the $c_5^A$-root could become negative
when reconstructed from the hierarchical fits. 
This problem seems to arise only when the fit of the multipole does not start from its own peak, 
\ie{} when $\dot{A}_0\neq 0$. 
While Eq.~\eqref{eq:templateA_c5} was successfully employed to describe the (2,2) multipole in 
Refs.~\cite{Albanesi:2021rby, Albanesi:2023bgi}, we find Eq.~\eqref{eq:templateA} to be more robust
for the fits of the (2,0) mode. 

\lettersection{Hierarchical global fits} 
Once that we have performed the primary fits on our dataset, we need to provide
global fits for the fitted coefficients $\lbrace c_2^A, c_3^A, c_2^\phi, c_3^\phi\rbrace$ 
over the parameter space, together with the needed numerical quantities,
$\lbrace A_0, \dot{A}_0, \omega_0 \rbrace$. As explained in the main text,
we start by performing a 1-dimensional fit over the test-mass simulations using 
a quadratic ansatz
\begin{equation}
f_{\rm \nu=0}(\hat{a}) = b_0 + b_1 \hat{a} + b_2 \hat{a}^2, 
\end{equation}
where $\hat{a}$ is the Kerr spin. The constant coefficient
is fixed to be the Schwarzschild value. 
Following the discussion of the main text on the phenomenologies of the (2,0) mode, 
for $A_0$ and $\dot{A}_0$  we impose $b_1=-b_0-b_2$ in order to have $f_{\nu=0}(1)=0$, and fit $b_2$
on the Kerr data. For the other quantities, we fit both $b_1$ and $b_2$.  
Note that for $\omega_0$ we only employ data up to $\hat{a}=0.6$ for this 1-dimensional fit,
since higher-degree polynomials would be needed to capture its behavior over the whole spin range $\hat{a}\in[-1,1]$
and, as already mentioned, the (2,0) mode is strongly suppressed for high positive spins. 
We then perform another 1-dimensional fit over $\nu$ for the non-spinning configurations,
\begin{equation}
f_{\tilde{a}_0=0}(\nu) = b_0 + c_1 \nu + c_2 \nu^2. 
\end{equation}
We then combine these two fits in a 2-dimensional ansatz,
\begin{align}
f_{\rm 2 D}(\ta, \nu) = & \left. b_0 + b_1    \ta + b_2    \ta^2  \right.     + \cr
                        & \left( c_1 + d_{11} \ta + d_{21} \ta^2\right) \nu   + \cr  
                        & \left( c_2 + d_{12} \ta + d_{22} \ta^2\right) \nu^2,
\label{eq:gfit2d}
\end{align}
where we remind that $\ta\equiv (\chi_1 m_1 + \chi_2 m_2)/M $ generalizes the Kerr spin $\hat{a}$, and the $d_{ij}$ coefficients
have to be determined from the data. For $A_0$ and $\dot{A}_0$, we impose $d_{2j} = -c_j-d_{1j}$, 
so that $f_{\rm 2 D}(1, \nu) = 0$, and we determine $d_{1j}$
using the remaining simulations of our dataset. The same simulations are used
to determine all the $d_{ij}$ coefficients for the other fitted quantities. 
The coefficients found with this approach are reported in Table~\ref{tab:gfit}
for all the needed quantities.

\begin{table*}[t]
	\caption{\label{tab:gfit} Coefficients of the hierarchical global fits performed with the ansatz of Eq.~\eqref{eq:gfit2d}
	over the test-mass simulations listed in Table~\ref{tab:sims_testmass} and waveforms generated with \Surr{}.
	}
\begin{center}
\begin{ruledtabular}
\begin{tabular}{c | c | c  c | c c | c c c c} 
 & $b_0$ & $b_1$ & $b_2$ & $c_1$ & $c_2$ & $d_{11}$ & $d_{21}$ & $d_{12}$ & $d_{22}$ \\
\hline
\hline
$A_0$       &  0.07647 & -0.13512 &  0.05865 &  -0.19955 &   0.33426 &   0.37027 &  -0.17072 &  -0.20921 & -0.12505 \\
$\dot{A}_0$ &  0.00786 & -0.01705 &  0.00919 &  -0.02534 &   0.02541 &   0.05440 &  -0.02905 &  -0.01746 & -0.00795 \\
$\omega_0$  &  0.08950 & -0.02150 & -0.00671 &   0.13470 &  -0.03056 &  -0.02966 &  -0.21276 &  -0.07630 &  0.71567 \\
$c_2^A$     &  0.13295 & -0.01104 & -0.02368 &  -0.08291 &   1.30014 &   0.11119 &   0.48070 &   0.61649 & -1.69515 \\
$c_3^A$     & -1.61888 & -0.20029 & -0.08509 &   2.52896 &   2.43339 &   3.61367 &   2.34373 & -10.88689 & -6.28732 \\
$c_2^\phi$  &  0.15313 &  0.00082 & -0.01702 &  -0.04679 &   1.23738 &  -0.08273 &  -0.01566 &   0.61189 & -0.03258 \\
$c_3^\phi$  &  4.41377 &  6.15871 &  5.23689 & -23.3638  & 240.998   & -45.5214  & -33.2851  & 347.687   & 185.324 \\

\end{tabular}
\end{ruledtabular}
\end{center}
\end{table*}

\lettersection{NQC and ringdown matching} 
In order to achieve an \ac{IMR} model, each multipole of the inspiral/plunge EOB waveform $h^{\rm inspl}_\lm$
has to be completed with a post-merger model $h^{\rm rng}_\lm$.
The matching between the two waveforms is performed as
\begin{equation}
\label{eq:eob_full}
h_\lm = h^{\rm inspl}_\lm \hat{h}_\lm^{\rm NQC} \theta(\tmatch-t) + h^{\rm rng}_\lm \theta(t-\tmatch),
\end{equation}
where $\theta(t-\tmatch)$ is the Heaviside step-function, $\tmatch$ is the matching time,
and the \ac{NQC} corrections $\hat{h}_\lm^{\rm NQC}$ ensures the continuity 
between the inspiral and ringdown waveforms. 
In state-of-the-art EOB models, the quasi-circular inspiral waveform $h_\lm^{\rm inspl}$ of the $m>0$ 
modes is factorized and resummed according to Ref.~\cite{Damour:2008gu}.
Further resummation procedures for the residual PN amplitude and phase corrections 
have been explored during the years. 
However, in this work we focus on the $m=0$ modes, and we only consider the formally Newtonian prescription 
outlined in Ref.~\cite{Chiaramello:2020ehz}, which yields Eq.~\eqref{eq:h20N} for the (2,0) mode. 
The \ac{NQC} corrections are explicitly written as
\begin{equation}
\label{eq:hnqc}
\hat{h}^{\rm NQC}_\lm = \left( 1 + \sum_{i=1}^{3}  a_i^\lm n_i \right) \exp{ \left( i  \sum_{j=1}^{3} b_j^\lm n_{j+3} \right) },
\end{equation}
where $n_i$ are functions written in terms of time-derivatives of the radius and
in terms of the conjugate momentum of the tortoise coordinate $p_{r_*}\equiv \sqrt{A/B} p_r$,
where $A$ and $B$ are the EOB metric potentials; see, \eg{}, Ref.~\cite{Damour:2014sva} for more details. 
A common choice for the higher modes is
\begin{subequations}
\begin{align}
n_1 & = \frac{p_{r_*}^2}{r^2\Omega^2}, \\
n_2 & = \frac{\ddot{r}}{r \Omega^2}, \\
n_3 & = n_1 \, p_{r_*}^2, \\
n_4 & = \frac{p_{r_*}}{r \Omega}, \\
n_5 & = n_4 \, \Omega, \\
n_6 & = n_5 \, p_{r_*}^2,
\end{align}
\end{subequations}
where $\Omega$ is the orbital frequency. 
The coefficients $a_i$ and $b_j$ of Eq.~\eqref{eq:hnqc} are determined 
by solving, in general, the following linear system
\begin{subequations}
\label{eq:nqc_syst}
\begin{align}
       A_\lm^{\rm EOB}(\tNQC)      & =            A_\lm^{\rm rng}(\tNQC),  \\
      \dot{A}_\lm^{\rm EOB}(\tNQC) & =       \dot{A}_\lm^{\rm rng}(\tNQC),  \\
     \ddot{A}_\lm^{\rm EOB}(\tNQC) & =      \ddot{A}_\lm^{\rm rng}(\tNQC),  \\
       \omega_\lm^{\rm EOB}(\tNQC) & =        \omega_\lm^{\rm rng}(\tNQC),  \\
 \dot{\omega}_\lm^{\rm EOB}(\tNQC) & =  \dot{\omega}_\lm^{\rm rng}(\tNQC),  \\
\ddot{\omega}_\lm^{\rm EOB}(\tNQC) & = \ddot{\omega}_\lm^{\rm rng}(\tNQC).
\end{align}
\end{subequations}
In this work we do not consider second-order derivatives for the (2,0) mode, and therefore include only
$\lbrace n_1, n_2, n_4, n_5\rbrace$ in the NQC base.
The quantities on the left-hand-side are evaluated from $h_\lm^{\rm inspl} \hat{h}_\lm^{\rm NQC}$,
while the right-hand-side is computed from the ringdown waveform. 
Note that the coefficients are found by solving the system~\eqref{eq:nqc_syst} at 
a specific time, $\tNQC$, but the NQC corrections are applied on an extended time interval.
Therefore, phase and amplitude should be ideally monotonic up to the matching time
for optimal results. 

\begin{figure}[t]
  \centering 
    \includegraphics[width=0.48\textwidth]{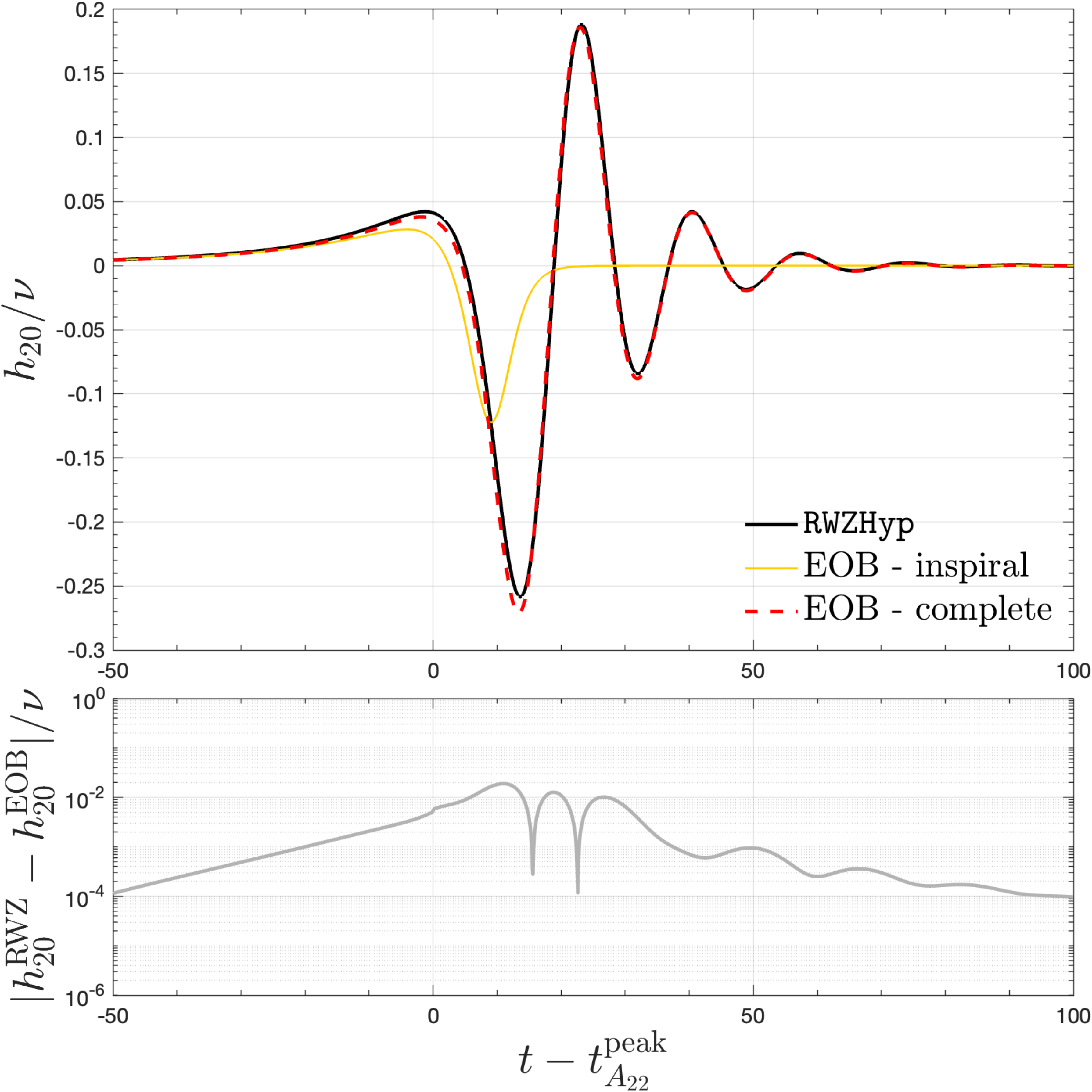}
    \caption{Comparison between the numerical and complete EOB waveform for 
    the Schwarzschild case (black and dashed red, respectively). 
    We also show the EOB inspiral waveform $h_{20}^{\rm inspl}$ (solid yellow).
    The residual is reported in bottom panel.}
  \label{fig:kerrorbits}
\end{figure}
\begin{figure*}[t]
  \centering 
    \includegraphics[width=0.32\textwidth]{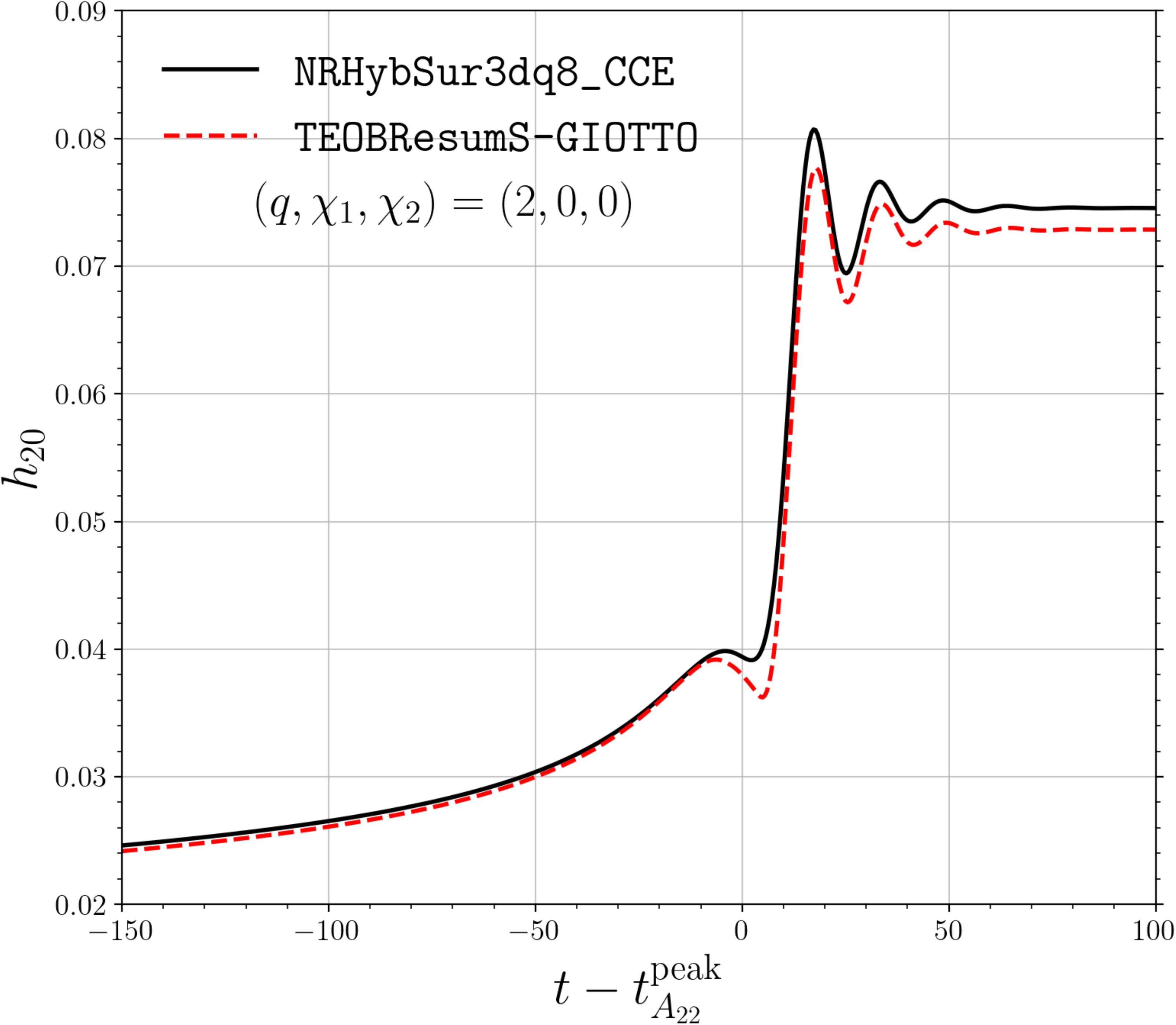}
    \includegraphics[width=0.32\textwidth]{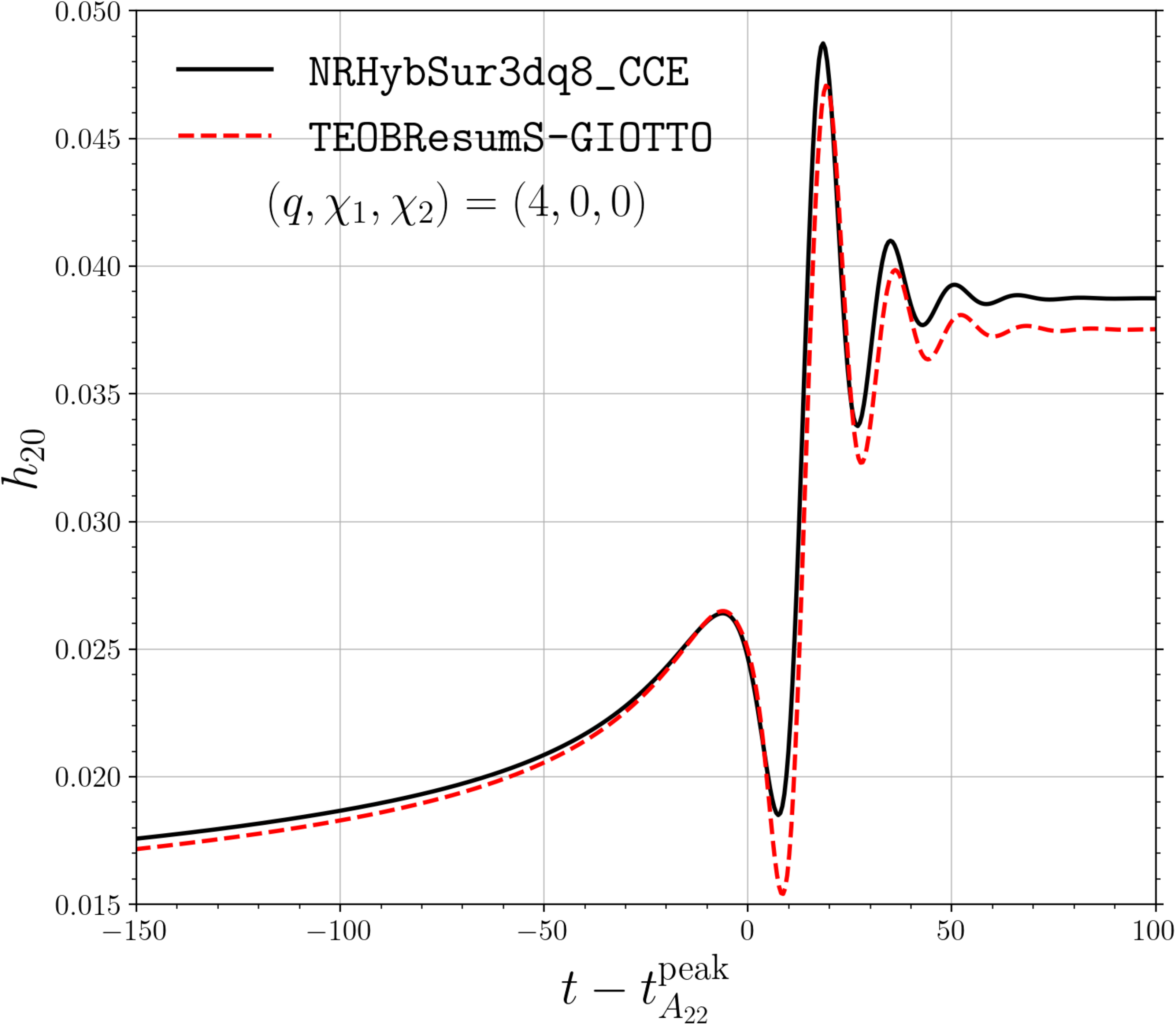}
    \includegraphics[width=0.32\textwidth]{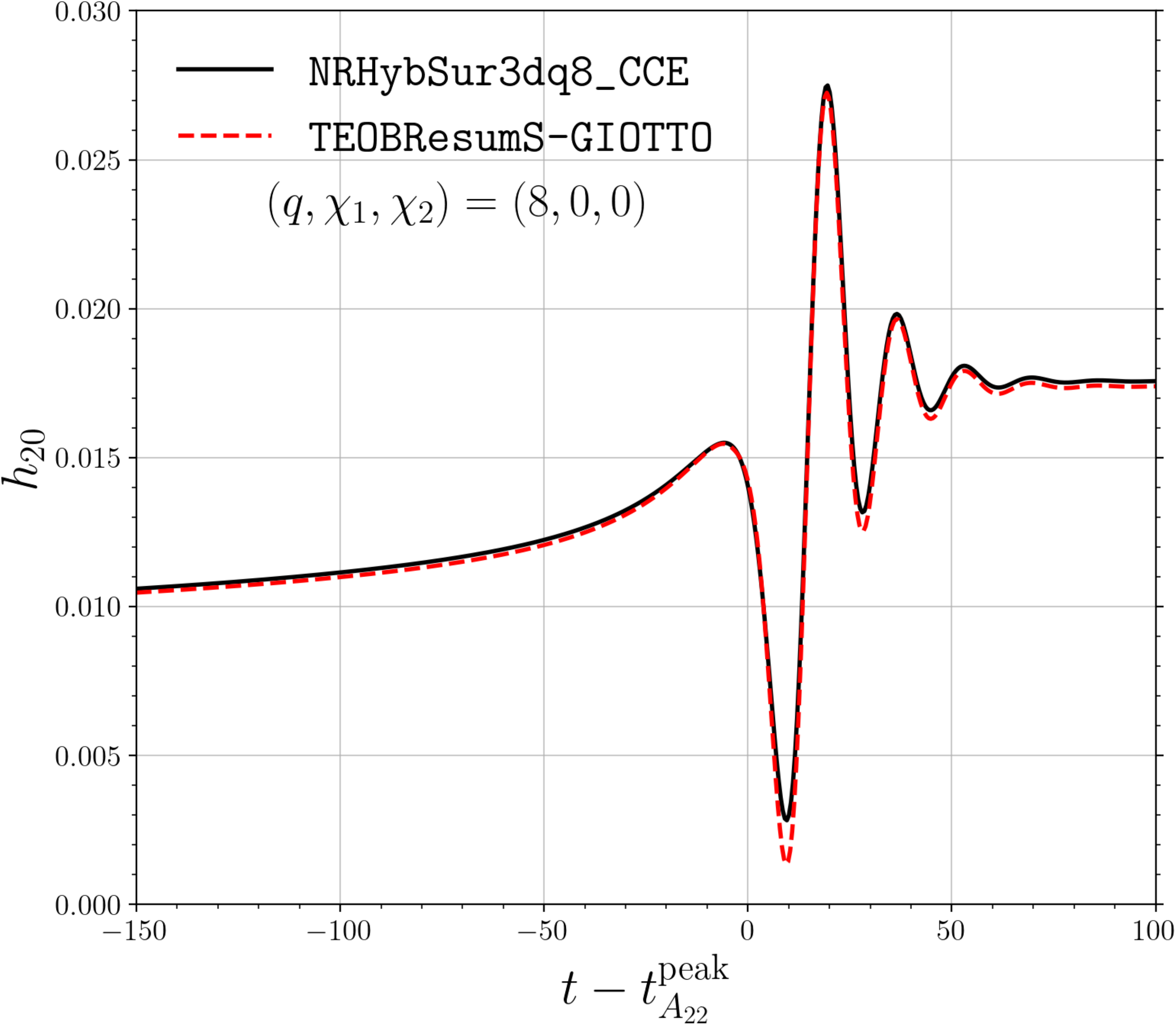}
    \includegraphics[width=0.32\textwidth]{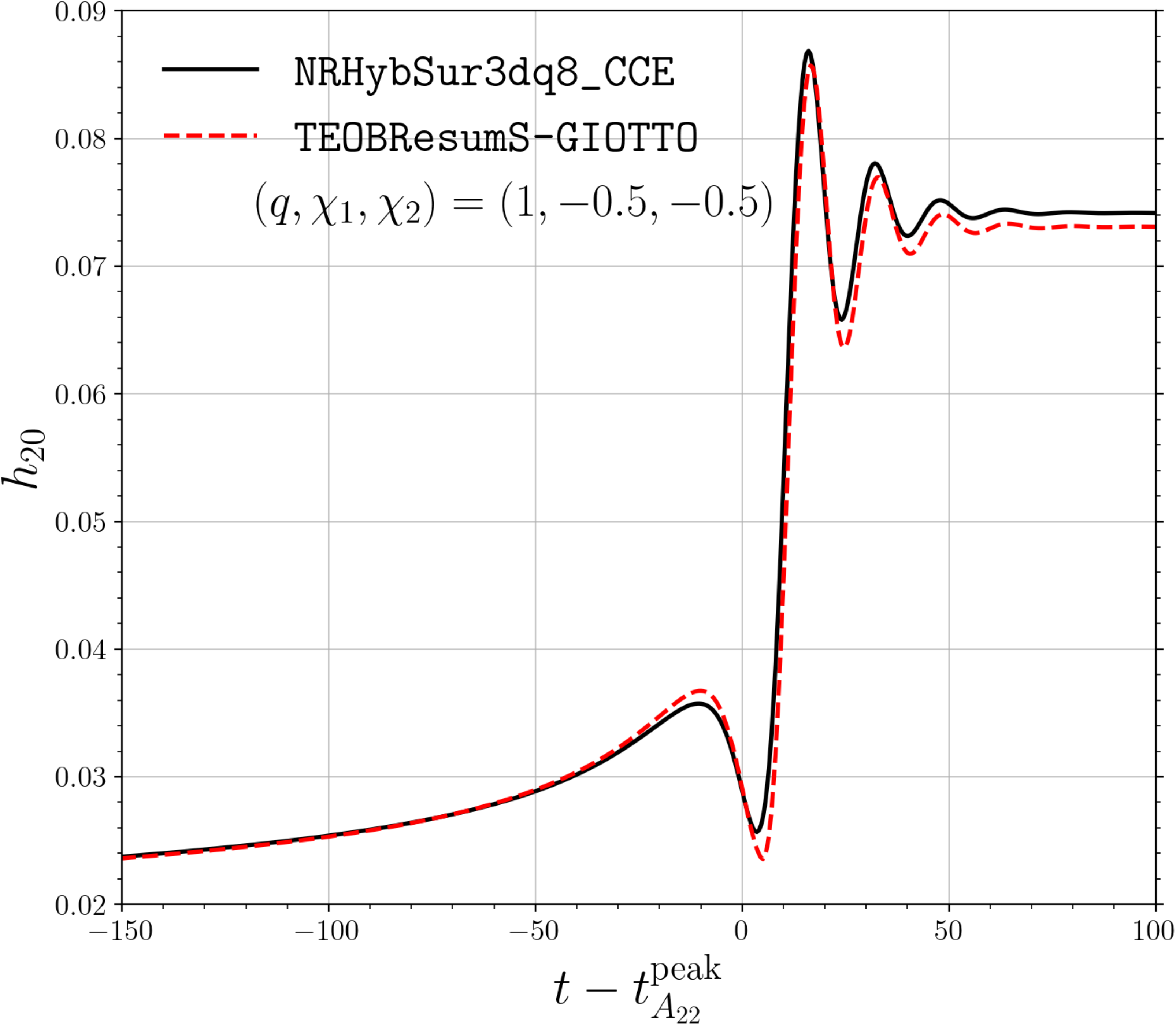}
    \includegraphics[width=0.32\textwidth]{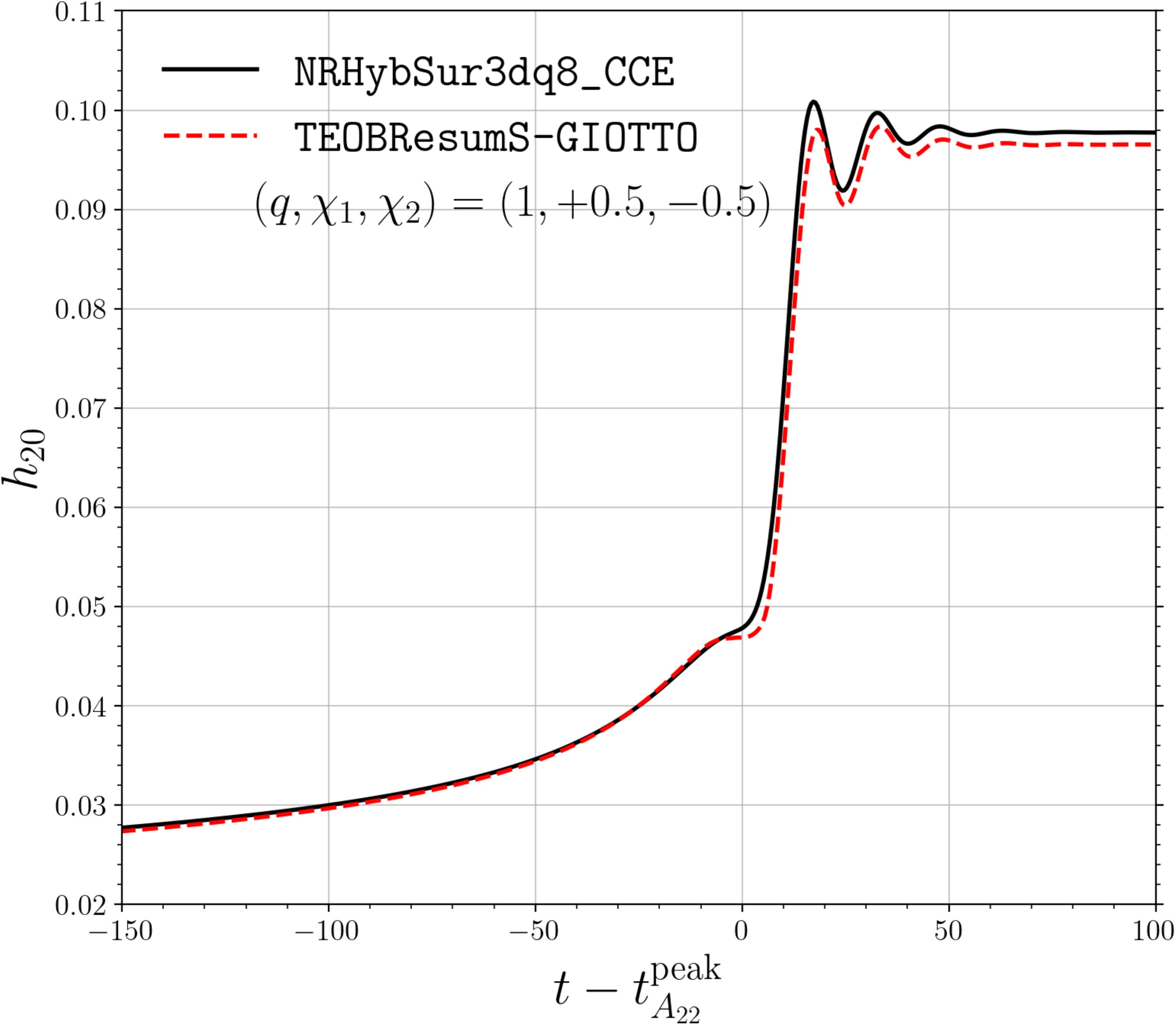}
    \includegraphics[width=0.32\textwidth]{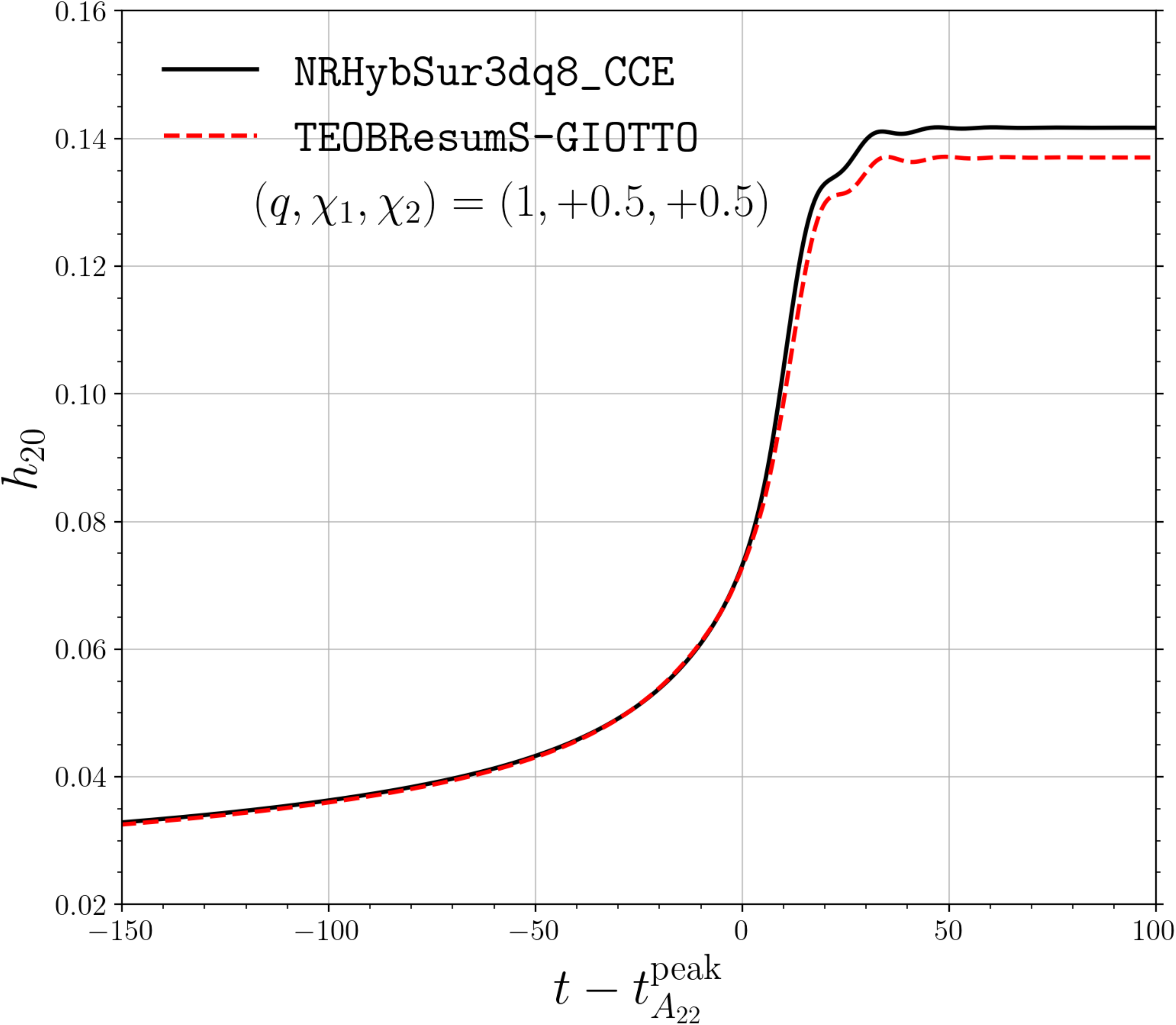}
    \includegraphics[width=0.32\textwidth]{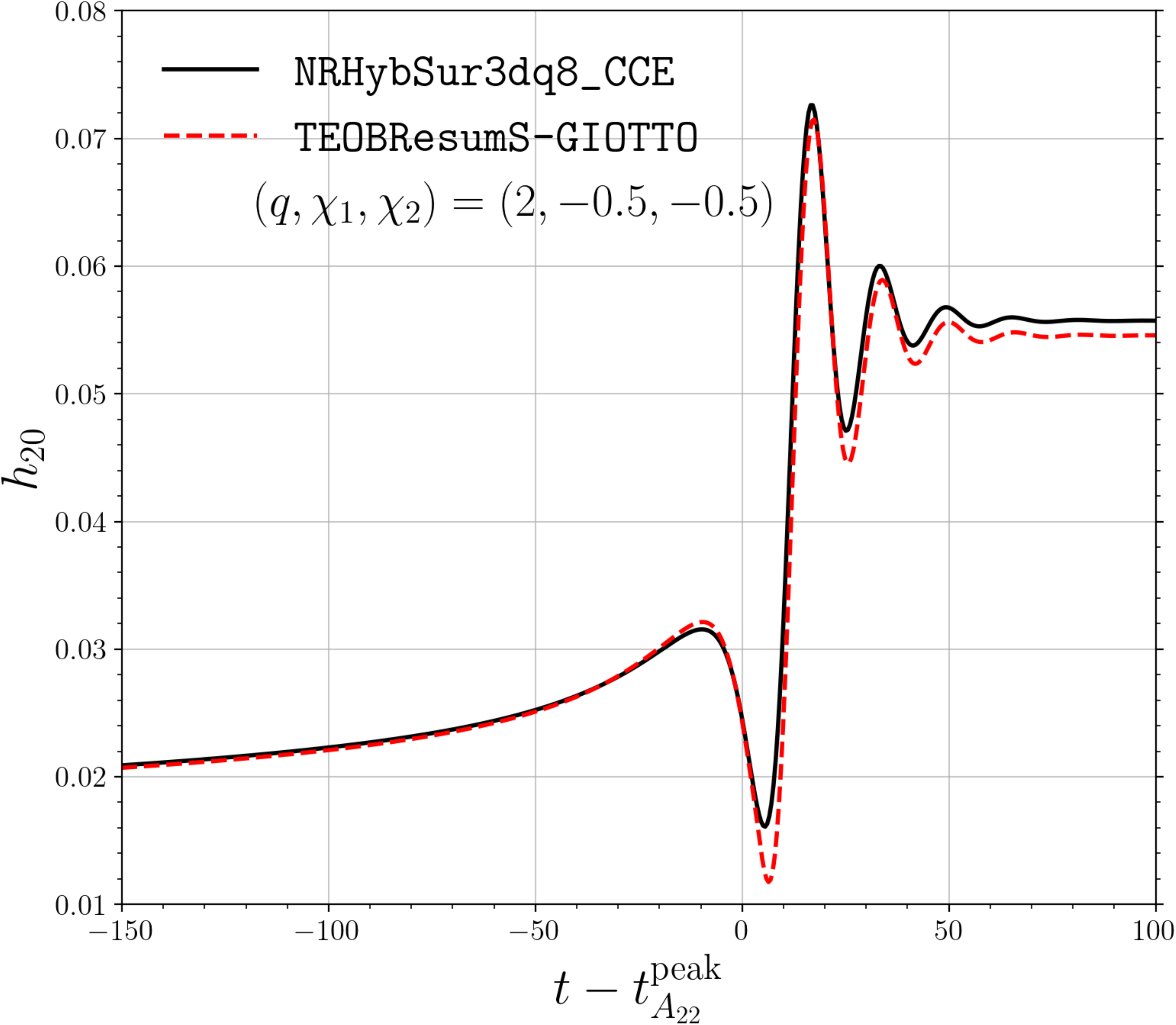}
    \includegraphics[width=0.32\textwidth]{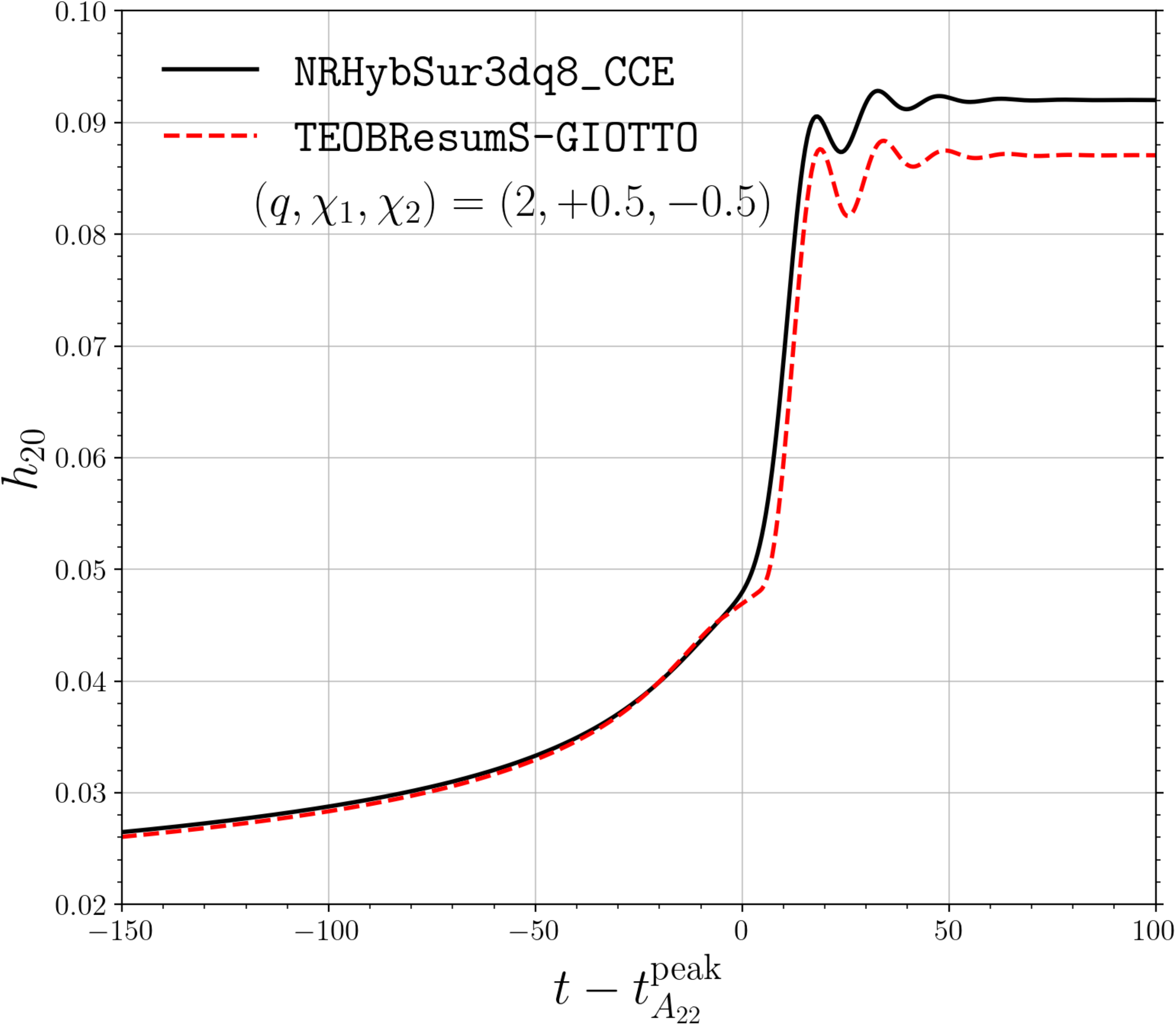}
    \includegraphics[width=0.32\textwidth]{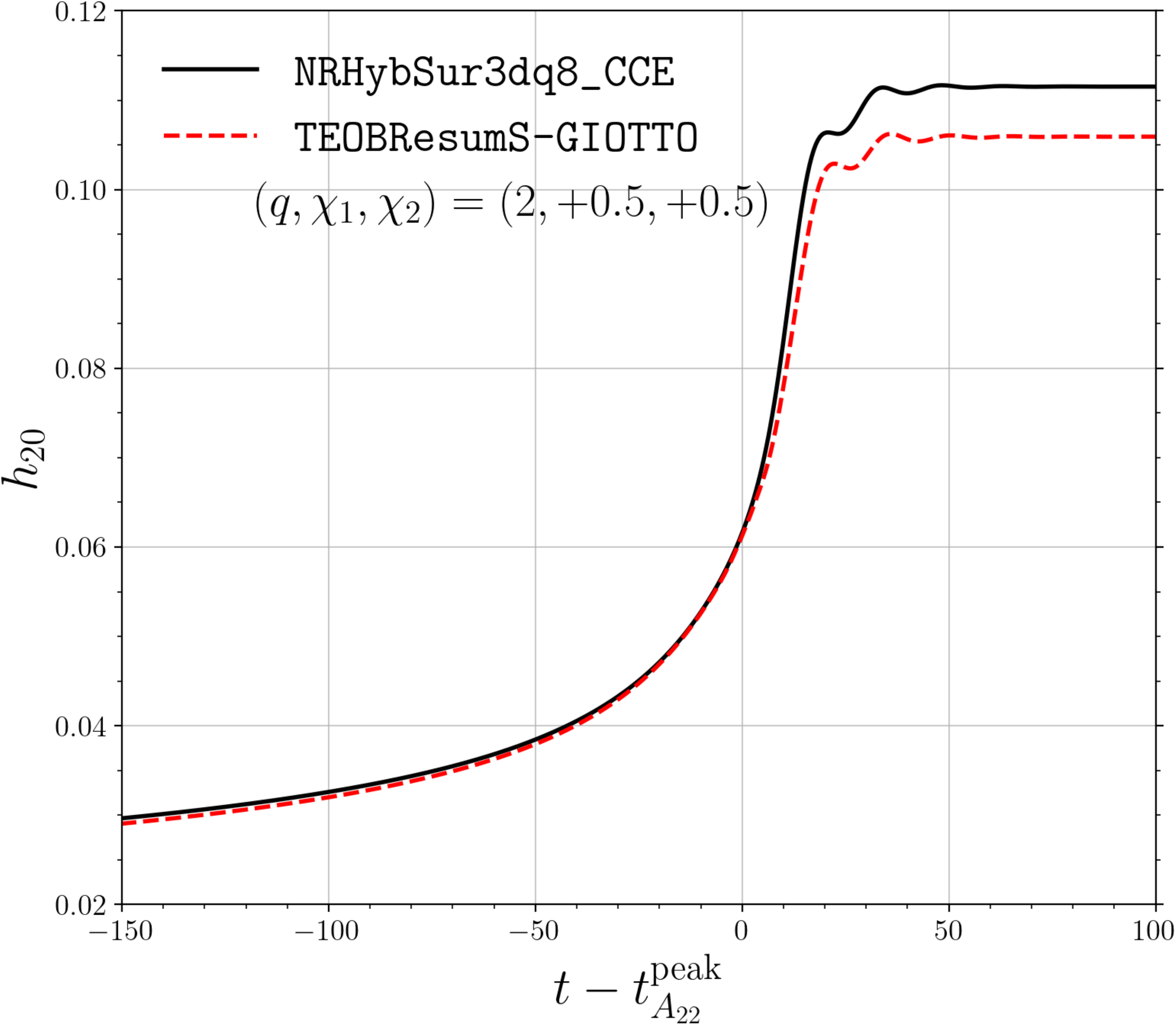}
    \caption{Comparisons of (2,0) modes for comparable mass configurations. We consider
    the modes from \SurrCCE{} and the ones computed in this work using \GIOTTO{} as a baseline.}
  \label{fig:eob_vs_cce_suppl}
\end{figure*}

By complexifying the oscillatory contribution of the (2,0) mode (or any other $m=0$ mode) via a Hilbert transform,
we obtain amplitudes and phases that are qualitatively similar to the ones of the $m>0$ modes, and we
can thus apply the NQC corrections to match the complexified inspiral (2,0) waveform
with the post-merger Hilbert model. We report in Fig.~\ref{fig:kerrorbits} the result
of this matching for the Schwarzschild case, where we used the most natural
choice $t_{20}^{\rm NQC}=\tA22$, which we tested to work well also
for the Kerr cases.
Note that the value of the EOB waveform at
the matching time is not equal to the
numerical one, despite the hierarchical fits
exactly incorporating the Schwarzschild limit. This is due to the fact 
that when the matching is performed, we rotate the complex ringdown signal 
of a certain phase determined by the inspiral/plunge waveform.
We thus lose the aforementioned property,
which is instead observed in the primary fits. 
For generic mass ratios, we find that the choice 
$t_{20}^{\rm NQC}=\tA22-2$ yields a better numerical/analytical agreement. 
In these cases, the right-hand-side of Eqs.~\eqref{eq:nqc_syst} 
is still evaluated from the post-merger waveform, which is extrapolated back in time up to $t_{20}^{\rm NQC}$. 
Finally, since we determine $a_i$ and $b_j$ at $t_{20}^{\rm NQC}=\tA22-2$ but 
still perform the match at $\tA22$, a small discontinuity would be observed at the 
matching time. This can be easily cured with a cubic patch applied between $t_{20}^{\rm NQC}-1$ and $\tA22+4$.
The four coefficients of the patch are determined by requiring $C_1$-continuity conditions at both ends. 
Note that this second patch is applied on the real physical mode after matching. 
This patch ensures a continue and reliable waveform, as shown from the EOB/NR comparisons
reported in Fig.~\ref{fig:eob_vs_cce} and~\ref{fig:eob_vs_cce_suppl}. 

\lettersection{Low-frequency PN-matching for null memory} 
Since we start our EOB evolution at finite radius and compute 
the (2,0) memory contribution from the corresponding fluxes
through Eq.~\eqref{eq:memo20}, we have $h_{20}^{\rm memo}\neq 0$ for 
$\Omega \rightarrow 0$. To cure this unphysical property, the result from Eq.~\eqref{eq:memo20} 
can be matched to the PN analytical formula, which has the correct low-frequency
behavior. We perform the matching with the 
3.5PN nonspinning result~\cite{Favata:2008yd,Cunningham:2024dog}
plus the 1.5PN spinning contribution~\cite{Cao:2016maz}. The complete expression thus reads
\begin{widetext}
\begin{align}
h_{20}^{\rm memo, 3.5PN} &= \frac{4}{7}\sqrt{\frac{5 \pi}{6}} \nu x \left\{ 1 + \left(-\frac{4075}{4032} + \frac{67}{48}\nu  \right) x +  \left( \frac{-2813+756\nu}{2400} \ta  +  \frac{3187}{2400} \delta \chi_A \right)x^{3/2} + \right. \cr
& \left(-\frac{151877213}{67060224} - \frac{123815}{44352} \nu + \frac{205}{352} \nu^2\right) x^2  +  \pi \left(-\frac{253}{336} +  \frac{253}{84}\nu\right)  x^{5/2}   + \cr
&  \left[ -\frac{4397711103307}{532580106240} +\left(\frac{700464542023}{13948526592} - \frac{205}{96}\pi^2\right)\nu+
\frac{69527951}{166053888}\nu^2  +  \frac{1321981}{5930496}\nu^3 \right] x^3 \cr 
&\left. \pi  \left(\frac{38351671}{28740096} - \frac{3486041}{598752}\nu - \frac{652889}{598752}\nu^2\right)  x^{7/2} \right\}
\end{align} 
\end{widetext}
where $\delta=(m_1-m_2)/M$, $\chi_A  = (m_1 \chi_1 - m_2 \chi_2)/M$, and $x=\Omega^{2/3}$. 
We then compute the residual between the memory contribution
obtained from the radiated energy flux and the 3.5PN expression, which is then
fitted with the ansatz ${\rm Res}(x) = d_0 + d_5 x^5 + d_{5.5} x^{11/2}$ for $t<t_{\rm end}^{\rm fit}$, where 
$t_{\rm end}^{\rm fit} = \tA22 - 500$. For the non spinning equal mass case, this time corresponds to $x\sim0.105$.  
The coefficient $d_0$ is then added to the result obtained from Eq.~\eqref{eq:memo20},
so that $h_{20}^{\rm memo} \rightarrow 0$ for $t\rightarrow-\infty$. 
Note that a similar approach has been followed in Ref.~\cite{Rossello-Sastre:2024zlr},
but with different choices for the amount of analytical PN information employed and, consequently, for the fitting ansatz of the residual.
We tested that choosing
different time intervals for the fit of the residual does not significantly influence $d_0$. 
Similarly, using less analytical information or more free coefficients in the ansatz yield equivalent results.
However, the PN series does not typically converge in strong field regimes, therefore one should
be careful in choosing $t_{\rm end}^{\rm fit}$ too close to the merger time.

\lettersection{EOB/NR unfaithfulness}
%
To assess the accuracy of the model more systematically across the parameter space, 
we consider frequency-domain mismatches between the full EOB and NR (2,0) multipoles. 
We compute the mismatch (or {\it unfaithfulness}) between two signals as 
\begin{equation}
  \bar{\mathcal{F}} = 1 - \max_{t_0, \phi_0} \frac{\langle h_1,h_2\rangle }{ \langle h_1,h_1\rangle  \langle h_2,h_2\rangle}\, ,
\end{equation}
where
\begin{equation}
  \langle h_i, h_j \rangle = 4 \Re\int_{f_{\rm min}}^{f_{\rm max}} \frac{\tilde{h}_i \tilde{h}_j^*}{S_n(f)} df\, ,
\end{equation}
and $\tilde{h}_i$ is the Fourier transform of $h_i$, while $S_n(f)$ is
the zero-detuned, high-power noise spectral density of Advanced LIGO~\cite{aLIGODesign_PSD}.
We make use of the package \texttt{pycbc}~\cite{Biwer:2018osg}.
With this noise, mismatches are usually computed in the frequency range $\left[ f_{\rm min}, f_{\rm max} \right] = \left[ 10,1024 \right]$ Hz, which we also
consider in this work.
Therefore, our computation mainly tests the accuracy of the oscillatory contribution of the waveform, which is the main 
novelty of this work. The memory, which is a low-frequency component, marginally influences this computation. 
Moreover, since the computation involves a numerical fast Fourier transform, ensuring periodicity is crucial to prevent spectral leakage. 
To address this in the case of signals with displacement memory, recent studies have explored different methodologies~\cite{Chen:2024ieh,Valencia:2024zhi}. 
In this work, we employ a time-domain windowing technique similar to that discussed in Sec.~IV~E.1 of Ref.~\cite{Chen:2024ieh}.

Mismatches calculated using only the (2,0) modes for binaries with mass ratio up to $q=8$ and different spins configurations 
with $\chi_i \in [-0.8,0.8]$ are reported in the right panel of Fig.~\ref{fig:eob_vs_cce} and discussed in the main text. 

Finally, when considering more multipoles as in the case of the non-spinning equal-mass quadrupolar waveforms discussed in the conclusions, 
we consider the sky-maximized overlap statistic~\cite{Harry:2017weg}. We have verified that, 
for current detections, the impact of the (2,0) mode on the mismatch of the $\ell=2$ quadrupolar waveforms is negligible. 
%

\lettersection{Additional time-domain comparisons} 
%
We report in Fig.~\ref{fig:eob_vs_cce_suppl} some additional time-domain EOB/NR comparisons between the complete (2,0) obtained with the method 
outlined in this work by using \GIOTTO{} as a baseline and the (2,0) mode of the surrogate model 
\SurrCCE{}. 
Visually, the main source of disagreement between the two model
is given by the different final offset, while the oscillatory parts are more similar. 
Considering that the offset of the modes is consistent before merger,
and considering that for the equal mass nonspinning case the final offset is essentially the same 
(cfr. with Fig.~\ref{fig:eob_vs_cce}),
we are prone to think that the source of this disagreement has to be searched in the amplitude
of the higher modes with $m>0$ during the merger-ringdown phase, as already argued in the
main text. The improvement of these modes is beyond the scope of this work, and we thus defer this investigation. 

\begin{table*}[t]
	\caption{\label{tab:sims_testmass} Quasi-circular test-mass simulations considered in this work.  
	We considered $\nu=10^{-3}$ in the dissipative part of the dynamics, but no $\nu$-corrections have been
	included in the Schwarzschild and Kerr Hamiltonians.}
\begin{center}
\begin{ruledtabular}
\begin{tabular}{c | c  c | c c c | c c | c c} 
code & $\hat{a}$ & $r_0$ & $A_0$ & $\dot{A}_0$ & $\omega_0$ & $c_2^A$ & $c_3^A$ & $c_2^\phi$ & $c_3^\phi$  \\
\hline
\hline
                  \Teuk{} &-0.90 &  9.35 & 0.2456169 & 0.0310443 & 0.1041740 &  0.12475& -1.48859&  0.14000&  2.70793 \\
                  \Teuk{} & -0.60 &  8.55 & 0.1776497 & 0.0212031 & 0.0995899 &  0.13025& -1.54394&  0.14473&  2.79897 \\
                  \Teuk{} & -0.50 &  8.25 & 0.1575544 & 0.0184084 & 0.0978831 &  0.13173& -1.55649&  0.14782&  3.03224 \\
                  \Teuk{} & -0.30 &  7.70 & 0.1222918 & 0.0136646 & 0.0950989 &  0.13344& -1.57392&  0.15165&  3.47555 \\
                  \RWZ{}  &  0.00 &  7.00 & 0.0764704 & 0.0078630 & 0.0895022 &  0.13295& -1.61888&  0.15313&  4.41377 \\
                  \Teuk{} &  0.30 &  5.90 & 0.0419396 & 0.0038958 & 0.0852170 &  0.12913& -1.64958&  0.15469&  6.06861 \\
                  \Teuk{} &  0.50 &  5.25 & 0.0225565 & 0.0018838 & 0.0779949 &  0.12231& -1.72724&  0.15075&  8.51613 \\
                  \Teuk{} &  0.60 &  4.90 & 0.0145450 & 0.0011256 & 0.0720827 &  0.11605& -1.80273&  0.14477& 10.69553 \\
                  \Teuk{} &  0.70 &  4.50 & 0.0080635 & 0.0005633 & 0.0640581 &   $\cdots$&   $\cdots$&   $\cdots$&   $\cdots$ \\
                  \Teuk{} &  0.80 &  4.10 & 0.0032148 & 0.0001903 & 0.0507208 &   $\cdots$&   $\cdots$&   $\cdots$&   $\cdots$ \\
                  \Teuk{} &  0.90 &  3.55 & 0.0004914 & 0.0000202 & 0.0267972 &   $\cdots$&   $\cdots$&   $\cdots$&   $\cdots$ \\
                  \Teuk{} &  0.95 &  3.20 & 0.0000337 & 0.0000008 & 0.0074474 &   $\cdots$&   $\cdots$&   $\cdots$&   $\cdots$ \\

\end{tabular}
\end{ruledtabular}
\end{center}
\end{table*}

\end{document}